\documentclass[11pt]{article}
\usepackage[margin=1.2in]{geometry}
\usepackage{enumitem}
\usepackage{amsmath}
\usepackage{listings}
\usepackage{hyperref}
\usepackage{graphicx}
\usepackage{hhline}
\usepackage{times}
\usepackage{bm}
\usepackage{afterpage}
\usepackage{comment}
\usepackage{amssymb}
\usepackage{indentfirst}
\usepackage[table]{xcolor}
\usepackage{xcolor}
\usepackage{verbatim}

\usepackage{tabularx,booktabs}
\usepackage{caption}
\usepackage[skip=3ex]{subcaption}

\usepackage{algorithm,algcompatible,amsmath}
\usepackage{setspace}

\usepackage[backend=bibtex, style=bwl-FU, maxcitenames=2]{biblatex}
\usepackage[font=small,textfont=it]{caption}
\algnewcommand\BLOCK{\item[\textbf{Block 1.}]}%
\algnewcommand\BLOCKK{\item[\textbf{Block 2.}]}%
\algnewcommand\BLOCKKK{\item[\textbf{Block 3.}]}%
\algnewcommand\BLOCKKKK{\item[\textbf{Block 4.}]}%

\algnewcommand\STEPA{\item[\textbf{Step (a)}]}%
\algnewcommand\STEPB{\item[\textbf{Step (b)}]}%
\algnewcommand\STEPC{\item[\textbf{Step (c)}]}%

\title{Dynamic Mixture of Finite Mixtures of Factor Analysers with Automatic Inference on the Number of Clusters and Factors}

\date{\today}

\author{Margarita Grushanina\thanks{Institute for Quantitative Economics, Vienna University of Economics and Business, Welthandelsplatz 1, 1020 Vienna, Austria. Email: \href{mailto:margarita.grushanina@s.wu.ac.at}{margarita.grushanina@s.wu.ac.at}} \and  Sylvia Fr{\"u}hwirth-Schnatter\thanks{Institute for Statistics and Mathematics, Vienna University of Economics and Business, Welthandelsplatz 1, 1020 Vienna, Austria. Email: \href{mailto:sfruehwi@wu.ac.at}{sfruehwi@wu.ac.at}}}

\newcommand{\diag}{\ensuremath{\mathrm{diag}}}

\newcommand{\alphaB}{\alpha_{\mathcal{B}}}
\newcommand{\alphaC}{\alpha_{\mathcal{C}}}
\newcommand{\alphaM}{\alpha_{\mathcal{M}}}
\newcommand{\Hkk}{H_k}

\begin{document}
\maketitle

\begin{abstract}
\noindent

  Mixtures of factor analysers (MFA) models represent a popular tool for finding structure in data, particularly high-dimensional data. While in most applications the number of clusters, and especially the number of latent factors within clusters, is mostly fixed in advance, in the recent literature models with automatic inference on both the number of clusters and latent factors have been introduced. The automatic inference is usually done by assigning a nonparametric prior and allowing the number of clusters and factors to potentially go to infinity. The MCMC estimation is performed via an adaptive algorithm, in which the parameters associated with the redundant factors are discarded as the chain moves.
While this approach has clear advantages, it also bears some significant drawbacks. Running a separate factor-analytical model for each cluster involves matrices of changing dimensions, which can make the model and programming somewhat cumbersome. In addition, discarding the parameters associated with the redundant factors could lead to a bias in estimating cluster covariance matrices. At last, identification remains problematic for infinite factor models.
The current work contributes to the MFA literature by providing for the automatic inference on the number of clusters and the number of cluster-specific factors while keeping both cluster and factor dimensions finite. This allows us to avoid many of the aforementioned drawbacks of the infinite models. For the automatic inference on the cluster structure, we employ the dynamic mixture of finite mixtures (MFM) model. Automatic inference on cluster-specific factors is performed by assigning an exchangeable shrinkage process (ESP) prior to the columns of the factor loading matrices. The performance of the model is demonstrated on several benchmark data sets as well as real data applications.


\end{abstract}

\section{Introduction}

Mixtures of factor analysers (MFA) models combine both clustering and local dimensionality reduction performed separately in each cluster and are particularly useful for modelling data with complex and nonhomogeneous structure. First works involving MFA models appeared already in the 1990ies when \cite{GH1996} developed an expectation-maximization (EM) algorithm for inference on the parameters of an MFA model. \cite{GB2000} later considered a Bayesian treatment of MFA via a variational approximation. At the same time, \cite{Fokoue2000} provided an "exact estimation" via an MCMC algorithm for inference on the MFA model, which was further ameliorated in \cite{FT2003}.
The following years have seen a fair amount of literature on various versions of MFAs. The most notable include \cite{MCNM2008}, who assessed an MFA model in the context of parsimonious Gaussian mixture models, and \cite{Viroli2010}, who introduced a mixture of factor mixture analysers (MFMA). The key feature of the MFMA model is that it assumes that the data are generated according to several factor models with a certain prior probability (thus performing a local dimension reduction at the first level), and that in each factor model the factors are described by a multivariate mixture of Gaussians (thus performing a global dimension reduction at the second level). 

Determining the number of clusters and the number of cluster-specific factors has always been a challenging issue.  Many authors either treat both as known or fixed, or, like, e.g. \cite{MCNM2008} and \cite{Viroli2010}, run their models for various number of components in the mixture and factors in the factor analytical part and use model selection criteria to choose the best fitting model. In an early attempt to find a way to learn the model dimensions from data, \cite{FT2003} developed a Birth-and-Death MCMC algorithm (which can be seen as a limiting case of the reversible jump MCMC) which uses the fact that the posterior distributions of both factor models and finite mixture models are invariant to permutations of the order of their parameters and thus the collection of their parameters can be viewed as a point process. More recently, \cite{Papastamoulis2018} introduced an overfitting Bayesian MFA (BMFA), which estimates the unknown number of mixture components assuming a fixed number of factors. The optimal number of factors is then determined using information criteria.

The most flexible BMFA model up to date, of which we are aware, is the infinite mixture of infinite factor analysers (IMIFA) model introduced in \cite{Murphy2019}, which allows an automatic inference on both the numbers of clusters and cluster-specific factors by assigning nonparametric priors to both cluster weights and cluster-specific factor loadings. For the automatic inference on the number of clusters they employ a Pitman-Yor process prior, using its stick-breaking representation and a slice sampler for MCMC estimation. Automatic inference on the cluster-specific number of factors is achieved with the multiplicative gamma process (MGP) prior of \cite{BD2011} and an adaptive Gibbs sampler is used to facilitate estimation with varying dimensions.

While the infinite MFA model has the clear advantage of eliminating the need to predefine the model dimensions, this flexibility comes with certain costs. Changing dimensions of cluster-specific factor models make the model somewhat cumbersome and challenging for efficient programming. Also, most of such methods employ adaptive MCMC algorithms, where all the parameters associated with the redundant factors are discarded at each iteration when adaptation occurs. As the discarded columns of factor loading matrices are though small but usually not exactly zero, some information is thus lost, which might introduce a bias in computing cluster-specific covariance matrices. At last, identification of the cluster-specific factor models remains an open issue for the infinite MFA models as there is no formal guarantee that the variance identification condition of \cite{Anderson1956}, which imposes an upper limit on the number of factors, is satisfied.

In this paper we suggest 
an innovative way to specify a Bayesian MFA model, which allows for the automatic inference on the number of clusters and cluster-specific factors.
This is achieved by exploring a finite-dimensional representation of the 
infinite nonparametric priors. For the mixture part, we employ the dynamic mixture of finite mixtures (MFM) model, introduced in \cite{TS2020}. 
This model puts a prior on the number of mixture components and allows inference with respect to the number of filled components which can be regarded as the number of clusters in the data. For each  
cluster-specific factor model, we generalize the cumulative shrinkage process (CUSP) prior introduced by \cite{LDD2020} and employ an exchangeable shrinkage process (ESP) prior. As shown by \cite{SFS2022}, such  a prior implicitly shrinks the factor loadings toward zero as the column index increases and allows inference on the number of active columns which can be regarded as the cluster-specific  number of factors. 

The rest of the paper is organised as follows. Section \ref{BMFA} introduces a general notion of a Bayesian MFA model. The IMIFA model is described in Section \ref{IMIFA-model}, and Section \ref{CUSP-prior} provides the details of the (CUSP) prior on factor loadings. Section \ref{Finite} describes the main idea of this paper, with the details of the dynamic MFM model for the clustering solution given in Section \ref{DynMFM} and the ESP prior on factor loadings explained in Section \ref{ESP}. The full specification of other priors and the choice of hyperparameters are described in Section \ref{hyper}. Section \ref{MCMC} presents our four block MCMC algorithm, based on the telescoping sampler introduced in \cite{TS2020}. The performance of our method is illustrated in an extensive simulation study in Section \ref{Simulation} and compared with several benchmark MFA models using publicly available data sets as well as some real data in Section \ref{Applications}. The paper concludes in Section \ref{Conclusion}.

\section{Bayesian MFA} \label{BMFA}
	
	In this section we provide a brief review of Bayesian MFA models. 
 Given $T$ observations $\bm{y} = (\bm{y}_1, \ldots, \bm{y}_T)$ of a multivariate $p$-dimensional random variable, a BMFA model is formulated as follows. It is assumed that these observations can be grouped into $K$ groups (clusters) and within these $K$ clusters, labelled by $k = 1, \ldots, K$, the variability of our observations can be described by $H_k$ latent factors. Note that the number of latent factors $H_k$ can vary across clusters.  
 To enable identification of the cluster-specific factor models, the number of latent factors in each cluster $H_k$ should satisfy the variance identification condition of \cite{Anderson1956}, which states that whenever an arbitrary row is deleted from the factor loading matrix, two disjoint matrices of rank $H_k$ remain. This imposes the upper bound on the number of factors of $H_k \leq \frac{p - 1}{2}$, which, however, is not always taken into account in the literature. 
 
 The model can be formalised in the following way.
 The variation of the observations within each cluster $k$ is described by a cluster-specific factor-analytical model:
\begin{align*} 
	\bm{y}_t - \bm{\mu}_k = \bm{\Lambda}_k \bm{f}_t^k + \bm{\epsilon}_t, 
\end{align*}
where $\bm{\mu}_k$ is a $p$-dimensional vector of cluster-specific means, $\bm{\Lambda}_k$ is a $p \times H_k$-dimensional cluster-specific factor loading matrix, $\bm{f}^k_t$ is a $H_k$-dimensional vector of latent factors, and $\bm{\epsilon}_t$ is a $p$-dimensional vector of idiosyncratic errors with cluster-specific variances.

It is usually assumed that the latent factors are orthogonal, namely:
\begin{align} \label{Factors}
\bm{f}_t^k \sim N_{H_k}(\bm{0},\bm{I}_{H_k}).
\end{align}
Furthermore, it is usually assumed that $\bm{f}^k_t$, $\bm{f}^k_s$, $\bm{\epsilon}_t$ and $\bm{\epsilon}_s$ are pairwise independent for all $t \neq s$.
Another important assumption concerns the distribution of the idiosyncratic errors $\bm{\epsilon}_t$:
\begin{align} \label{Errors}
\bm{\epsilon}_t \sim N_p(\bm{0}, \bm{\Xi}_k), \quad \quad \bm{\Xi}_k = \diag(\xi^2_{1k}, \ldots, \xi^2_{pk}).
\end{align}
The assumption (\ref{Errors}) implies that conditional on the common factors $\bm{f}^k_t$ all $p$ elements of $\bm{y}_t$ are independent, so all dependencies between the variables are explained by the common factors. The assumptions (\ref{Factors}) and (\ref{Errors}) imply that the data $\bm{y}$ can be considered arising from a multivariate normal distribution. Taking into account the clustered structure of the data, for each observation $t$ independently, we can formulate the following mixture model:
\begin{align*}
f(\bm{y}_t \, | \, \bm{\mu}_k, \bm{\Lambda}_k, \bm{\Xi}_k) = \sum_{k=1}^K \eta_k N_p\left(\bm{y}_t; \bm{\mu}_k, \bm{\Omega}_k \right), \quad \quad \bm{\Omega}_k = \bm{\Lambda}_k \bm{\Lambda}_k^T + \bm{\Xi}_k,
\end{align*}
where $\bm{\Omega}_k$ denotes the cluster-specific covariance matrix of the data and $\eta_k, (k=1, \ldots, K)$ are cluster weights. Note that this decomposition of $\bm{\Omega}_k$ into the sum of the 
cross-covariance matrix $\bm{\Lambda}_k \bm{\Lambda}_k^T$ 
and the idiosyncratic errors' covariance matrix is possible only under the assumptions (\ref{Factors}) and (\ref{Errors}).

The main challenge usually lies in establishing the values of $K$ and $H_1, \ldots, H_K$. 
	
\subsection{IMIFA model} \label{IMIFA-model}

Although most of the MFA literature requires the values of $K$ and $H_k$ to be pre-specified, recently there have been two notable attempts to relax this restriction. \cite{Papastamoulis2018} suggest an overfitting BMFA model, where the number of "alive" clusters is derived using overfitting mixtures in the spirit of \cite{fru-mal:fro} and the number of factors is determined using information criteria. This model has been further extended in \cite{Papastamoulis2020} to a parsimonious MFA to include eight different parametrizations depending on whether various constraints are applied or not (\cite{MCNM2008}). However, this model only allows the same number of factors in all clusters, which imposes a rather strong restriction for a large variety of data.

Another way of avoiding the need to pre-specify the number of clusters and factors is to use a nonparametric prior and allow $K$ and all $H_k$s to go to infinity in theory. There is a relatively vast literature on nonparametric priors for factor models while substantially less attention has been given to the MFA type models. Recently, \cite{Murphy2019} developed an infinite mixture of infinite factor analysers (IMIFA) model, which assigns nonparametric priors to both the number of clusters $K$ and the number of cluster-specific factors $H_k$, thus providing for a fully automatic inference on the cluster and factor structure of the data and at the same time allowing the number of factors to differ between clusters.

To identify the number of clusters, \cite{Murphy2019} assign a Pitman-Yor process (PYP) prior to the mixture weights. This prior has a stick-breaking representation, which can be summarized as:

\begin{eqnarray} \label{IMIFA}
&& v_k \sim \mathcal{B}(1-d, \sigma + kd),  \qquad   \bm{\theta}_k \sim 
G_0, \\
&& \eta_k = v_k \prod_{l=1}^{k-1}(1-v_l),  \qquad  G 
= \sum_{k=1}^{\infty} \eta_k \delta_{\bm{\theta}_k} \sim PYP(\sigma, d, G_0),
\nonumber
\end{eqnarray}
where $\delta_{\bm{\theta}_k}$ is the Dirac measure
centred at $\bm{\theta}_k$, such that draws are composed of a sum of infinitely many point masses. The PYP reduces to the Dirichlet process (DP) when $d = 0$. Thus, in comparison to the DP, the PYP implies an a priori smaller number of components by shifting the mass to the left. Sampling from the PYP is performed via a slice sampler.

For inference on the number of factors in each cluster, \cite{Murphy2019} employ the multiplicative gamma process (MGP) prior introduced in \cite{BD2011} for the factor loadings $\lambda_{ihk}$ of the $k$th cluster factor loading matrix $\bm{\Lambda}_k$. This prior can be formulated as follows:
\begin{eqnarray} \label{MGP}
&& \lambda_{ihk} \rvert \phi_{ihk},\tau_{hk}, \sigma_k   \sim N(0, \phi_{ihk}^{-1} \tau_{hk}^{-1} \sigma_k^{-1}),  \qquad \phi_{ihk} \sim \mathcal{G}(\nu_1/2, \nu_2/2),   \qquad  \tau_{hk} = \displaystyle\prod_{l=1}^{h} \delta_{lk}, \nonumber \\
&& \delta_{1k} \sim \mathcal{G}(a_1, b_1),   \qquad  \delta_{lk} \sim \mathcal{G}(a_2, b_2),   \qquad  l\geq2, 
\end{eqnarray}
where $\delta_{lk}$ $(l = 1, \ldots, \infty)$ are independent, $\tau_{hk}$ is a column shrinkage parameter for the $h$-th column of the $k$-th cluster loading matrix $\bm{\Lambda}_k$, $\forall$ $k = 1, \ldots, \infty$. The $\tau_{hk}$s are stochastically increasing under the restriction $a_2 > 1$, which favours growing shrinkage as the column index increases. $\phi_{ihk}$ 
are local shrinkage parameters for the elements of the $h$-th column of $\bm{\Lambda}_k$ and are designed to favour sparsity while also preserving non-zero loadings. Finally, $\sigma_k$ is the cluster shrinkage parameter, which reflects the belief that the degree of shrinkage is cluster-specific.

Initially, a conservative starting number of factors $H_0$ is chosen, which is supposed to be clearly bigger than any possible real number of factors. The adjustment of the factor dimensionality in each cluster $k$ is performed by means of an adaptive Gibbs sampler. This requires choosing some small neighbourhood of zero $\epsilon_0$. Then, at iteration $g$ of the sampler, if some chosen proportion of elements of the $h$-th column of the $k$-th cluster loading matrix  $\bm{\Lambda}_k$ is within this neighbourhood of zero $\epsilon_0$, factor $h$ is discarded with all corresponding parameters. If no factor can be discarded at iteration $g$, a new factor is generated and all the corresponding parameters are sampled from the prior distributions. This adaptation is performed at iteration $g$ with probability $p(g) = e^{\alpha_0 + \alpha_1 g}$, where the parameters $\alpha_0$ and $\alpha_1$ are chosen so as to satisfy the diminishing adaptation condition of \cite{Roberts2007}.

While this model represents an important benchmark in nonparametric factor models, it has some serious drawbacks. The hyperparameters $a_1$ and $a_2$ in (\ref{MGP}) control both the shrinkage rate and the prior for loadings on active factors, which creates a trade off between the need to maintain a rather diffuse prior for the active terms and shrinkage for the redundant ones. This leads to a problem, illustrated in \cite{Durante2017}, that the efficient shrinkage conditions imposed on the hyperparameters\footnote{\cite{Durante2017} showed that the condition $a_2 > 1$ is not sufficient for efficient shrinkage and two more conditions, namely, $a_2 > b_2 + 1$ and $a_2 > a_1$, are required.} provide too strong shrinkage in larger data sets.

\subsection{CUSP prior on factor loadings} \label{CUSP-prior}
	
\cite{LDD2020} proposed a nonparametric prior on the variances of the elements of the factor loading matrix, which largely corrects the drawbacks of the MGP prior. This prior and its properties were further studied and generalised in \cite{Kowal2021} in the context of nonparametric functional bases. This cumulative shrinkage process prior, which represents a sequence of spike-and-slab distributions, assigns a growing mass to the spike as the model complexity grows. Active loadings are controlled by the slab parameters, while inactive loadings are controlled by the spike parameters. Although this prior has not yet been implemented for MFA models it can easily be adjusted to the MFA framework as follows, assuming a given number $K$ of mixture components: 
\begin{eqnarray} \label{CUSP}
&& \lambda_{ihk}\,|\,\theta_{hk} \sim N(0, \theta_{hk}), \qquad \text{where} \:\, i = 1,\ldots, p, \:\,  \:\,  h = 1, \ldots, \infty, \:\, \text{and} \:\, k = 1,\ldots, K , \\
&& \theta_{hk}\,|\,\pi_{hk} \sim (1-\pi_{hk})p_{slab}(\theta_{hk} |\phi_{\theta}) + \pi_{hk} \delta_{\theta_{\infty}}, \quad \pi_{hk} = \sum_{l=1}^h w_{lk}, \quad w_{lk} = v_{lk} \prod_{m=1}^{l-1}(1-v_{mk}), \nonumber
\end{eqnarray}
where $\theta_{hk}$ is a column shrinkage parameter for the $h$th column of the cluster-specific factor loading matrix $\bm{\Lambda}_k$, $\pi_{hk} \in (0,1)$ is the probability of the spike, and the $v_{hk}$ are generated independently from $\mathcal{B}(1, \alphaC )$, 
following the usual DP stick-breaking representation (\cite{set:con}). $\phi_{\theta}$ denotes the hyperparameters of the slab distribution and $\delta_{\theta_{\infty}}$ is chosen in \cite{LDD2020} at 0.05. However, it can be replaced by some continuous distribution without affecting the key properties of the prior, as shown in \cite{Kowal2021}, where a normal mixture of inverse-gamma prior is employed for the spike and slab distributions. 

The adaptation of the factor dimensionality is performed differently than in \cite{BD2011}. While the probability of adaptation at iteration $g$ of the sampler is also set to satisfy the diminishing adaptation condition of \cite{Roberts2007}, there is no need to pre-specify the neighbourhood of $0$.  The "inactive"  columns of $\bm{\Lambda}_k$ are identified as those which are assigned to the spike and are discarded at iteration $g$ with probability $p(g) = e^{\alpha_0 + \alpha_1 g}$ together with all corresponding parameters. If at iteration $g$ all columns of the factor loading matrix in cluster $k$ are identified as "active", i.e. assigned to the slab, an additional column of $\bm{\Lambda}_k$ is generated from the spike and all the corresponding parameters are sampled from their prior distributions. The initial number of columns $H$, at which the CUSP model is truncated by assuming $v_{Hk}=1$, is set equal to $p+1$, as there can be at most $p$ active factors and by construction at least one column is assigned to the spike.

The assignment of the columns of the factor loading matrices to spike or slab at iteration $g$ is done using $H_k^{(g)}$ categorical variables $z_{hk} \in \{1,2, \ldots, H_k^{(g)}\}$ with a discrete prior $Pr(z_{hk} = h \, | \, w_{hk}) = w_{hk}$,
where $H_k^{(g)}$ is the number of columns in $\bm{\Lambda}_k$ at iteration $g$.
Given $z_{hk}$, the spike-and-slab prior (\ref{CUSP}) is represented as follows:
\begin{align} \label{eq5A}
\theta_{hk} \,|\, z_{hk} \sim I\{z_{hk} \leq h \} P_{spike}(\theta_{hk}) + (1 - I\{z_{hk} \leq h \}) P_{slab} (\theta_{hk}).
\end{align}
The number of active columns in the cluster-specific loading matrix $\bm{\Lambda}_k$ is then defined as 
$H_k^{* (g)} 
= \sum_{h=1}^{H_k^{(g)}} I \{z_{hk} > h\}$. 

\section{From infinite to finite} \label{Finite}

While the above mentioned nonparametric models have the important advantage of allowing fully automatic inference on the cluster and factor dimensionality of the data set, they also have significant drawbacks. Changing dimensions of a separate factor model in each cluster make the model cumbersome and programming somewhat complicated. For example, to run the $H_k$-factor model with the CUSP prior in each cluster in the MFA framework with $K$ clusters, one would require $\sum_{k=1}^K H_k^2$ density evaluations in classifications into spike or slab. Moreover, the discarded inactive columns of the loading matrices $\bm{\Lambda}_k$ have their elements close to but not exactly zero. This information loss might create a bias in estimating the cluster covariance matrices $\bm{\Omega}_k$. Also, as it has been shown in \cite{Schiavon2020}, the choice of the truncation criteria in the adaptation mechanism of the MCMC sampler in the infinite factor models is rather influential. This leads to more uncertainty when working with data sets where there is no clear indication of the probable number of latent factors. Furthermore, when it comes to identification of factor loadings, the theoretically infinite number of factors in the model can hinder variance identification.

In this section, we propose a new type of Bayesian MFA model with automatic inference on the dimensionality of both cluster and factor structure of the data, which, at the same time, allows to keep both $K$ and $H_k$ finite, while being random variables that are inferred from the data. To achieve this, we first employ a dynamic mixture of finite mixtures (MFM) model, introduced in \cite{TS2020}, to identify the cluster structure of the data. Using the MFM model gives an additional advantage of making the MCMC estimation possible solely within Gibbs sampler steps relying on telescoping sampling, eliminating the need to refer to additional methods such as slice sampling. For the cluster-specific factor-analytical part of the model we propose a finite representation of the CUSP prior, which belongs to the class of more general exchangeable shrinkage process (ESP) priors (\cite{SFS2022}). We call our novel MFA model dynamic Mixture of Finite Mixtures of Factor analysers ($(MF)^2 A$), where the term "dynamic" refers to applying the dynamic MFM for the clustering solution and reflects the fact that the concentration parameter of the Dirichlet prior on cluster weights depends on the number of components $K$, see the prior formulation in (\ref{weights}) below.

	\subsection{Dynamic mixture of finite mixtures of of factor analysers} \label{DynMFM}
	
Let $k = 1, \ldots, K$ denote the cluster index and $h = 1, \ldots, H$ denote the indices of factors within a cluster. The dynamic mixture of finite mixtures (MFM) model is mixture model with a prior on $K$ and can be written in the following hierarchical way (\cite{TS2020}):
\begin{align}
	K &\sim p(K), \nonumber\\ 
	\eta_1,\ldots,\eta_K\,|\,K, \alphaM
 &\sim Dir_K \left(\frac{\alphaM
 }{K} \right), \label{weights} \\ 
	\bm{\mu}_k \,| \,\bm{b}_0,\bm{B}_0 
 &\sim N_p
 (\bm{b}_0,\bm{B}_0), \quad \text{cluster means for}\: k = 1, \ldots,K ,\nonumber \\ 
	\bm{\Omega}_k &= \bm{\Lambda}_k \bm{\Lambda}^T_k + \bm{\Xi}_k, \quad \text{cluster covariance matrices for} \: k = 1, \ldots,K, \nonumber \\ 
	S_t \,| 
 \eta_1,\ldots,\eta_K &\sim \mathcal{M}(1;\eta_1,\ldots,\eta_K) ,\quad \text{latent allocation variables for}\: t = 1, \ldots,T , \nonumber \\ 
	\bm{y}_t \,| 
 S_t=k, \bm{\mu}_k, \bm{\Omega}_k &\sim N_p
 ( 
 \bm{\mu}_k, \bm{\Omega}_k) \quad \text{for each data point in} \: 1,\ldots,T. \nonumber
\end{align}
Under this model, the joint distribution of the data $\bm{y} = (\bm{y}_1,\ldots,\bm{y}_T)$ has a representation as a countably infinite MFM with $K$ components:
\begin{align*}
p( \bm{y}
) = \sum_{K=1}^\infty p(K) \prod_{t=1}^{T}\sum_{k=1}^K \eta_k N_p(\bm{y} ;
\bm{\mu}_k, \bm{\Omega}_k),
\end{align*}
where $p(K)$ is a prior for the number of mixture components. The model is called "dynamic" due to the fact that the Dirichlet concentration parameter $\frac{\alphaM
}{K}$ is inversely proportional to the number of components $K$, which favours more sparse solutions as the number of components grows.

In this framework, $K$ is defined as the (theoretical) number of components in the mixture, while the number clusters $K_+$ is defined as the number of filled components that generated the data, namely $K_+ = \sum^K_{k=1} \mathcal{I} \{|T_k| > 0\}$, where the set $T_k=\{t: S_t = k\}$  collects the indices of all observations generated by the component $k$ and the cardinality $|T_k|$ is the number of such observations.

Through the model, the latent allocation variables $\bm{S} = (S_1, \ldots, S_T)$ induce a random partition $\mathcal{C} = \{\mathcal{C}_1, \ldots, \mathcal{C}_{K_+} \}$ of the $T$ observations into $K_+$ clusters, where each cluster $\mathcal{C}_k$ contains all observations generated by the same mixture component (belonging to the same cluster). Assigning a prior to $K$ has the advantage that both $K$ and $K_+$ are random a priori. Depending on the choice of hyperparameters, they can be close or rather different, see the detailed investigation in \cite{gre-etal:spy}. 
 Having the Dirichlet concentration parameter decrease with increasing $K$ allows a gap between $K_+$ and $K$ and thus ensures randomness in the prior distribution of $K_+$ for a vast variety of different $\alphaM$ and $T$. Following the considerations in \cite{TS2020} and \cite{Grun2021}, we chose the suggested translated beta-negative-binomial (BNB) prior $K-1 \sim BNB(\alpha_\lambda, a_\pi, b_\pi)$, which represents a hierarchical generalisation of the Poisson, the geometric and the negative-binomial distributions. The p.m.f. takes the following form for $K=1, \ldots, \infty$:
\begin{align} \label{BNB}
p(K) = \frac{\Gamma(\alpha_\lambda + K - 1) B(\alpha_\lambda + a_\pi, K - 1 + b_\pi)}{\Gamma(\alpha_\lambda) \Gamma(K) B(a_\pi, b_\pi)}
\end{align}
where $\alpha_\lambda$, $a_\pi$ and $b_\pi$ are hyperparameters. The choice of hyperparameters can be governed by the desired value of the prior mean $E(K) = 1 + \alpha_\lambda \frac{b_\pi}{a_\pi - 1}$, which exists as long as $\alpha_\pi > 1$. An important advantage of this prior is that the three parameters $\alpha_\lambda$, $a_\pi$ and $b_\pi$ allow simultaneous control over both the expectation of $p(K)$ and its tails, as well as the implied prior on $K_+$ and its expectation (see \cite{TS2020} and \cite{gre-etal:spy} for details on the induced prior on $K_+$).

Since the hyperparameter $\alphaM$
in the Dirichlet concentration parameter $\frac{\alphaM}{K}$ plays an important role for the prior distribution induced on the number of filled clusters $K_+$ and the partitions, we adjust it to the data by assigning it a prior and updating it from the posterior distribution in a random walk Metropolis-Hastings step. We choose the F-distribution prior $\alphaM \sim \mathcal{F}(\nu_l,\nu_r)$ as it is flexible enough to allow various cluster solutions by modeling the behaviour close to zero and in the tail independently (see \cite{TS2020} for further motivation of this prior choice).

It is shown in \cite{TS2020}, that the dynamic MFMs can be regarded as a generalization of the Dirichlet process prior beyond the class of Gibbs-type priors. In fact, a Dirichlet process mixture (DPM) is the limiting case of a dynamic MFM where the prior $p(K)$ increasingly concentrates all prior mass at $K = +\infty$. Especially interesting is the connection of the dynamic MFMs to PYP mixtures. As shown in e.g. \cite{Blasi2015}, there exists a second family of PYP mixtures, where, using the notation in (\ref{IMIFA}), $\sigma < 0$ and $d = K |\sigma|$, with $K \in \mathbb{N}$ being a natural number. In the corresponding stick-breaking representation, stick $v_K = 1$ a.s., so this prior yields a mixture with infinitely many components, of which only $K$ have non-zero weights, with the symmetric Dirichlet distribution $\mathcal{D}_K (|\sigma|)$ acting as a prior. \cite{TS2020} show by comparison of the exchangeable partition probability functions (EPPF) that for dynamic MFMs, the prior partition distribution arises from mixing a PYP$(-\frac{\alphaM}{K}, \alphaM)$ prior over the reinforcement parameter $\sigma_K = -\frac{\alphaM}{K}$, while the concentration parameter $d = \alphaM$ is fixed.
	
\subsection{ESP prior for factor loadings} \label{ESP}

\cite{SFS2022} introduces the general class of exchangeable shrinkage process priors, which take the form of unordered spike-and-slab priors. Adjusted for the MFA framework, this prior is defined as follows. Let for each cluster $k = 1, \ldots, K$, assume $\bm{\tau}_k = \{ \tau_{hk} \in (0,1) \}, \, h = 1, \ldots, H$ to be a finite sequence of i.i.d. random parameters taking values in the unit interval. Let $\Theta_k = \{ \theta_{hk} \}, \, h=1, \ldots, H$ be a finite sequence of model parameters and assume that the $\theta_{hk} | \tau_{hk}$ are independent conditional on $\bm{\tau}_k$  and independent of all $\tau_{lk}, \, l \neq h$ for all $h$. If $p(\theta_{hk} | \tau_{hk})$ takes the following spike-and-slab form:
\begin{align} \label{ESPdef}
\theta_{hk} | \tau_{hk} \sim (1-\tau_{hk}) p_{spike}(\theta_{hk}) + \tau_{hk} p_{slab}(\theta_{hk}),
\end{align}
then $\Theta_k$ follows an exchangeable shrinkage process (ESP) prior. By definition, this prior is invariant to permutations of both the column and the cluster indices of $\theta_{hk}$, which makes it exchangeable.

It is often assumed in the literature that the slab probabilities $\tau_{1k}, \ldots, \tau_{Hk}$ follow a beta distribution, where the first parameter depends on $H$ see, e.g. \cite{SFSDHHL2022}, \cite{Rockova2016} amongst others. 
$H$ is here the same in all clusters and can be considered as the maximum possible number of factors, which the data allows. A typical choice of the beta prior for $\tau_{hk}$ would be 
\begin{align*} 
\tau_{hk} | H \sim \mathcal{B} \left( b_0 \frac{\alphaB
}{H}, b_0 \right), \quad h = 1, \ldots, H.
\end{align*}
This prior was proposed in \cite{SFSDHHL2022} in the context of sparse finite Bayesian factor models. For $H \rightarrow \infty$ it converges to the infinite two-parameter beta prior introduced by \cite{GGS2007} in the framework of Bayesian nonparametric latent feature models. With $b_0 =1$, this prior becomes the one-parameter beta prior employed by \cite{Rockova2016}:
\begin{align}  \label{1BP}
\tau_{hk} | H \sim \mathcal{B} \left( \frac{\alphaB 
}{H}, 1 \right), \quad h = 1, \ldots, H.
\end{align}
With $H \rightarrow \infty$ the prior (\ref{1BP}) converges to the Indian buffet process prior (see \cite{Teh2007} for more details). It is shown in \cite{SFS2022} that any ESP prior admits a finite generalised CUSP representation. 
In the context of our Bayesian MFA model, it is obtained by the permutation of the columns index $h$ of the
parameters $\theta_{1k}, \ldots, \theta_{Hk}$ according to the decreasing slab probabilities $\tau_{(1k)} > \ldots > \tau_{(Hk)}$ for each cluster $k$. Thus, the CUSP prior of \cite{LDD2020}, defined
in (\ref{CUSP}), can be considered as the limiting case of the following exchangeable spike-and-slab prior on $\theta_{hk}$ with $H<\infty$:
\begin{align} \label{taueq1}
\theta_{hk}\,|\,\tau_{hk} \sim \tau_{hk} p_{slab} (\theta_{hk}|\phi_\theta ) + (1-\tau_{hk}) p_{spike} (\theta_{hk}|\phi_0) , \quad \tau_{hk} \sim \mathcal{B}(\frac{\alphaB}{H},1), \quad h=1, \ldots, H ,
\end{align}
where $\phi_\theta$ and $\phi_0$ are the hyperparameters of the slab and the spike distributions, respectively. For $H \rightarrow \infty$, the hyperparameter $\alphaB$ coincides with the hyperparameter $\alphaC$ in (\ref{CUSP}).\footnote{However, while \cite{LDD2020} assume a fixed value $\alphaC = 5$, we adapt $\alphaB$ to data under a suitable prior, see Section \ref{hyper}.} 
Increasing spike probabilities $\pi_{hk}$ for  $h=1, \ldots, H$, as in the case of the CUSP prior, are obtained for each cluster from the decreasing order statistics $\tau_{(1k)}> \ldots > \tau_{(Hk)}$ by defining $\pi_{hk}=1 - \tau_{(hk)}$.

Representation (\ref{taueq1}) allows to choose an upper limit for the number of factors in each cluster, $H$, and keep it fixed throughout the model. By performing classification between spike and slab independently for each column, we will end up with defining an effective number of active factors $\Hkk$ 
in each cluster $k$, which is random both apriori and aposteriori, typically smaller than $H$, and varies across clusters. 

This relationship between (\ref{taueq1})  and the CUSP prior holds regardless of the distributions of the spike and the slab, both of which are allowed to depend on (random) hyperparameters $\phi_\theta$ in the slab and $\phi_0$ in the spike. Following \cite{LDD2020} and \cite{Kowal2021}, we combine the spike and slab distributions $p_{spike} (\theta_{hk}|\phi_0 )$ and $p_{slab} (\theta_{hk}|\phi_\theta )$ with a Gaussian scale mixture prior for the factor loadings in column $h$ of cluster $k$:
\begin{align*}
\lambda_{ihk}\,|\,\theta_{hk} \sim N(0, \theta_{hk}), \quad \text{where} \:\, i = 1,\ldots, p, \:\,  h = 1, \ldots, H \:\, \text{and} \:\, k = 1, \ldots, K.
\end{align*}
This allows to work out the marginal prior for the $h$th column $\bm{\lambda}_{hk}=(\lambda_{1hk}, \ldots, \lambda_{phk})^\top $
of the $k$th cluster factor loading matrix in case of 
suitably chosen distributions $p_{spike} (\theta_{hk}|\phi_0 )$ and $p_{slab} (\theta_{hk}|\phi_\theta)$. E.g., under the slab prior $ \theta_{hk} | a_\theta, b_\theta  \sim \mathcal{G}^{-1}\left(a_\theta, b_\theta \right) $, a Student-$t$ distribution results for $\bm{\lambda}_{hk}$, i.e. $\bm{\lambda}_{hk}| a_\theta, b_\theta  \sim t_{2 a_\theta}(\bm{0},
 b_\theta/a_\theta \bm{I}_p)$.
 
Thus, the full specification of the prior on the factor loadings in each cluster of the MFA model can be formalised as follows:
\begin{eqnarray} \label{MFA-Lambda}
&& \lambda_{ihk}\,|\,\theta_{hk} \sim N(0, \theta_{hk}), \quad \text{where} \:\, i = 1,\ldots, p, \:\,  h = 1, \ldots, H  \:\, \text{and} \:\,  k = 1, \ldots, K, \\ \nonumber
&& \theta_{hk} \,|\,\tau_{hk} \sim \tau_{hk} \mathcal{G}^{-1}(a_{\theta}, b_{\theta}) + (1-\tau_{hk}) \mathcal{G}^{-1}(a_0, b_0), \quad \quad \tau_{hk} \sim \mathcal{B}(\frac{\alphaB}{H},1).\\ \nonumber
\end{eqnarray}
By integrating out $\theta_{hk}$, each factor loading $\lambda_{ihk}$ gets the marginal prior
\begin{align*}
\lambda_{ihk} \sim \tau_{hk} t_{2a_\theta}(0, b_\theta/a_\theta) + (1-\tau_{hk}) t_{2a_0}(0, b_0/a_0),
\end{align*}
however, note that all loadings $\lambda_{ihk}$ within each column $h$ are dependent.

For MCMC estimation, we use the usual technique of data augmentation for ESP priors. We introduce $K H$ latent binary indicator variables $I_{hk}$, one for each column $h=1, \ldots,H$ of the loading matrix $\bm{\Lambda}_k$ in each cluster $k$, to classify the columns into \lq\lq active\rq\rq  
\ and \lq\lq inactive\rq\rq \ ones. The indicator $I_{hk}$ takes the value of either zero or one for each column $h = 1, \ldots, H$ and follows the Bernoulli prior $P(I_{hk} = 1 | \tau_{hk}) = \tau_{hk}$.

\subsection{Other priors and hyperparameters} \label{hyper}
	
We use the BNB prior on the number of components $K$ as in (\ref{BNB}), with the parameters $\alpha_\lambda = 1$, $\alpha_\pi = 4$ and $\beta_\pi = 3$, which results in the a priori expectation of the number of components $E(K)=2$. The reasoning behind this choice of hyperparameters can be found in \cite{TS2020} and \cite{Grun2021} along with a comparative study of the performance of various translated priors for $K-1$ in the MFMs context. 
For the hyperparameters $\nu_l$ and $\nu_r$ of the F distribution prior on the concentration parameter $\alphaM$,  
used in the prior for mixture component weights $\bm{\eta}_k = \{ \eta_k, \, k=1, \ldots, K \}$,
we chose $\alphaM
\sim \mathcal{F}(6,3)$ following the reasoning in \cite{TS2020}.

For the cluster means, we follow \cite{MW2016} and choose in (\ref{weights}) the independence prior
$\bm{\mu}_k \sim N_p(\bm{b}_0, \bm{B}_0)$ with the data-dependent hyperparameters
\begin{align*}
\bm{b}_0 = median(\bm{y}), \quad \bm{B}_0 = \diag(R_1^2, \ldots, R_p^2),
\end{align*}
where $R_i$ is the range of the data in dimension $i$.

In the application of mixture models to clustering multivariate data, it is often suggested in the literature to employ a hierarchical data-driven inverse Wishart prior for cluster covariance matrices (see, e.g. \cite{MW2016}, \cite{SFS2006}). In the MFA context, where each cluster contains a factor-analytical model, the cluster covariance matrices are computed at each iteration of the MCMC sampler as $\bm{\Omega}_k = \bm{\Lambda}_k \bm{\Lambda}_k^T + \bm{\Xi}_k$, where $\bm{\Lambda}_k$ is the $p \times H$ factor loading matrix of the cluster $k$ and $\bm{\Xi}_k$ is the $p \times p$ matrix of uniquenesses of the factor model in the cluster $k$. Thus, the prior on $\bm{\Omega}_k$ has a more general structure than an inverse Wishart prior and is driven by the prior choices for $\bm{\Lambda}_k$ and $\bm{\Xi}_k$.

The choice of the maximum possible number of factors $H$ is governed by the 
variance identification constraints. As mentioned in Section \ref{BMFA}, the variance identification of a factor model is guaranteed only when the number of latent factors satisfies the constraint $H_k \leq \frac{p-1}{2}$. Consequently, we set $H$ equal to the largest integer which is less or equal to $\frac{p-1}{2}$. We noticed, however, that in practical implementation in cases when the data dimensionality $p$ is rather small, like $p \leq 10$, which consequently leads to the upper limit on the number of factors being $\leq 5$, setting $H=p$ leads to better mixing and thus better performance of the model. Nevertheless, the effective number of active factors discovered by the model usually satisfies the identification constraint $H_k \leq \frac{p-1}{2}$.

The cluster-specific idiosyncratic variance parameters $\xi^2_{ik}$ are given a hierarchical prior:
\begin{align*}
\xi^2_{ik} \sim \mathcal{G}^{-1}(a_{\xi},b_{\xi i}), \quad \quad \quad b_{\xi i} \sim \mathcal{G}(a_g, b_{gi}).
\end{align*}
where the rate hyperparameters $b_{gi}$ are assigned the data-driven values $\frac{100}{R^2_i}$, following the considerations in \cite{StephensPhD1997} and \cite{SFS2006}. Assigning a data-driven hierarchical prior to $\xi^2_{ik}$ is particularly beneficial in the context of the MFA models due to the specific structure of the cluster-specific covariance matrices $\bm{\Omega}_k = \bm{\Lambda}_k \bm{\Lambda}_k^T + \bm{\Xi}_k$, where the cluster-specific idiosyncratic covariance matrix $\bm{\Xi}_k$ represents an important part of the covariance structure specific to cluster $k$. With the priors for the elements of $\bm{\Lambda}_k$ containing no data-related information, the prior for $\xi^2_{ik}$ provides $\bm{\Omega}_k$ with the link to the information from the data. This is especially important because, as explained in details in Section \ref{MCMC}, this prior is used to fill the parameters of the newly generated empty clusters during MCMC sampling, which makes the prior choice highly influential for the performance of the algorithm.

With the prior on factor loadings described in Section \ref{ESP}, the parameters of the spike and the slab distributions deserve some closer attention. As these parameters (especially of the spike) are rather influential in classifying factors into \lq\lq active\rq\rq\ and \lq\lq inactive\rq\rq\ ones (see, for example, \cite{Schiavon2020} for a discussion of this subject) we let them be determined by data by assigning hyperpriors to the scale parameters of both the spike and the slab as in (\ref{MFA-Lambda}), namely $b_0$ and $b_\theta$. To the scale parameter of the slab distribution $b_{\theta}$ we assign a gamma hyperprior $b_{\theta} \sim \mathcal{G}(a_2, b_2)$, with the hyperparameters $a_2$ and $b_2$ chosen such as to allow a rather flat distribution. With the spike distribution we aim at setting the hyperparameters in such a way, that the variance of $\theta_{hk}$
around zero at the spike is a priory expected at some small number, for example $0.05$ as in \cite{LDD2020}, but at the same time has enough flexibility to be bigger or smaller depending on the data. The mean of the inverse gamma distribution with parameters $\alpha$ and $\beta$ is defined as $\frac{\beta}{\alpha-1}$, which leads to the condition $\frac{b_0}{\alpha_0-1} = 0.05$. For $a_0$ reasonably big (the reason why this is a reasonable choice for our model is explained below), we need to choose the parameters of the hyperprior on $b_0$ in such a way that the mean of $\theta_{hk}$ at the spike is approximately $0.05$, which is easily done with the gamma hyperprior $b_0 \sim \mathcal{G}(a_1, b_1)$.

The choices of the spike and slab shape parameters $a_0$ and $a_{\theta}$ have to guarantee that the regions where the spike distribution dominates the slab distribution are centered around 0, while the  slab  distribution  dominates the spike distribution in the tails. A necessary condition for that is that the degrees of freedom parameter  $a_0$ in the spike is considerably larger than the degrees of freedom parameter  $a_\theta$ in the slab. 

Finally, following the idea to learn all the influential parameters from data, we assign a gamma hyperprior to the strength parameter of the beta prior for the slab probabilities $\alphaB \sim \mathcal{G}(a_\alpha, b_\alpha)$. Our specific choices of hyperparameters are summarized in Table~\ref{tab:hyperpar}.

\renewcommand{\arraystretch}{1.1}
\begin{table}[t!]
  \begin{center}
    \normalsize
    \begin{tabular}{c|c|c} 
    \hline
      Parameter(s) & Hyperparameter(s) & Value(s) \\
      \hline
      $\bm{\mu}_k$ & $\bm{b}_0, \bm{B}_0$ & $median(\bm{y}),diag(R^2_1, \ldots, R^2_p)$ \\
      $K$ & $\alpha_\lambda, \alpha_\pi, \beta_\pi$ & $1,4,3$ \\
      $\alpha$ & $\nu_l, \nu_r$ & $6,3$ \\
      $\alphaB $ & $a_\alpha, b_\alpha$ & $6,2$ \\
      $\xi^2_{ik}$ & $a_\xi, a_g, b_{gi}$ & $1, 3, 100/R^2_i$ \\
      $\theta_{hk}| I_{hk} =1$  & $a_\theta, a_2, b_2$ & $3,2,1$ \\
      $\theta_{hk}| I_{hk} =0$  
      & $a_0, a_1, b_1$ & $21,1,1$ \\
      \hline
    \end{tabular}
  \end{center}
  \vspace{-5mm}
  \caption{Hyperparameter specifications for the $(MF)^2 A$ model.}
  \label{tab:hyperpar}
\end{table}

\section{Posterior computations and MCMC algorithm} \label{MCMC}

\subsection{Nested Gibbs sampler} \label{Gibbs}

Despite the relatively complex nature of the model, with separate factor-analytical models nested within a cluster structure, posterior inference can be done solely within Gibbs sampler steps without referring to additional means, such as, for example, slice sampler in the case of using the PYP prior on cluster weights.

The sampler consists of four major blocks, where in the first block the partition is updated and $K_+$ non-empty clusters are identified. In the second one, the factor model algorithm is performed for every filled cluster and the number of active factors $H_k$ in each cluster $k = 1, \ldots, K_+$ is identified via the non-zero elements in the corresponding  columns of the binary indicator matrix $\bm{I} = \{I_{hk}\}$, where by $\bm{I}_k$ we denote the row of $\bm{I}$ which corresponds to cluster $k$. In the same way, $\bm{\theta}_k = \{ \theta_{hk}, \, h = 1, \ldots, H \}$ denotes the variances of the factor loadings in the cluster $k$. In the third block, the new number of mixture components $K \geq K_+$ is sampled and the Dirichlet parameter $\alphaM$ is updated via a random walk Metropolis-Hastings step. Finally, in the fourth block, we fill the empty clusters including the parameters of the underlying factor models from the corresponding priors. Thus, the first and the third blocks are the standard telescoping sampler clustering steps, as described in \cite{TS2020}. The full details of the sampler are provided in Algorithm \ref{Alg1}.

 \begin{algorithm}[t!]
    \caption{Telescoping sampling for the dynamic $(MF)^2 A$ model} \label{Alg1}
    \small
  \begin{algorithmic}[1]
    
    \vspace{0.2cm}
    \BLOCK 
    \item[(a)] Update the partition $\mathcal{C}$ of the data by sampling latent allocation indicators $S_t$ for $t = 1, \ldots, T$, from $Pr(S_t=k \, | \, \bm{\eta}_K, \bm{\mu}_1, \ldots, \bm{\mu}_K, \bm{\Omega}_1, \ldots, \bm{\Omega}_K, K) \propto \eta_k N_p(\bm{y}_t ;
    \bm{\mu}_k, \bm{\Omega}_k), k = 1, \ldots, K$.

	\item[(b)] Compute the number of observation points in each cluster $|T_k| = \#\{t \, | \, S_t=k\}$, the number of non-empty components $K_+ = \sum_{k=1}^{K} \mathcal{I} \{|T_k| > 0\}$, and relabel the components so that the first $K_+$ clusters are non-empty.
    
    \vspace{0.2cm}
    \BLOCKK
   	\item[(a)] For each of the filled clusters $ k = 1, \ldots, K_+$ run the factor analytical procedure with the spike-and-slab prior on factor loadings, sequentially updating $\bm{f}_t^k$, $\bm{\Lambda}_k$, $\bm{\Xi}_k$, $\bm{\theta}_k$, $\bm{I}_k$ and $\bm{\tau}_k$. Conditional on $\bm{\Lambda}_k$ and $\bm{\Xi}_k$ update cluster means $\bm{\mu}_k$ for the filled clusters $ k = 1, \ldots, K_+$.
   	
   	\item[(b)] Update the hyperparameters of the factor analytical model, conditional on $K_+$, $\bm{\theta}_k$, $\bm{I}_k$, $\bm{\Xi}_k$ and $\bm{\tau}_k$.
   	
   	\vspace{0.2cm}
	\BLOCKKK
	\item[(a)] Conditional on the partition $\mathcal{C}$, draw a new value of $K \geq K_+$ from 
\begin{align*}
p(K \, | \, \mathcal{C}, \alphaM) 
\propto p(K) \frac{\alphaM^{K_+} K!}{K^{K_+}(K-K_+)!}\prod_{k=1}^{K_+}\frac{\Gamma(|T_k| 
+ \frac{\alphaM}{K})}{\Gamma(1 + \frac{\alphaM}{K})}, \qquad K = K_+, K_++1, K_++2, \ldots .
\end{align*}

	\item[(b)] Using a random walk MH step, sample $\alphaM \,|\, \mathcal{C}, K$ from
\begin{align*}
p(\alphaM \,|\, \mathcal{C},K) \propto 
p(\alphaM)\frac{\alphaM^{K_+} 
\Gamma( \alphaM)}
{\Gamma(T + \alphaM)}\prod_{k=1}^{K_+}\frac{\Gamma(|T_k| + \frac{\alphaM}{K})}{\Gamma(1 + \frac{\alphaM}{K})}.
\end{align*}

	\vspace{0.2cm}
	\BLOCKKKK
	\item[(a)] If $K > K_+$, add $K-K_+$ empty clusters and sample their cluster means $\bm{\mu}_k$ and the corresponding factor model parameters, i.e. $\bm{\Lambda}_k$ and $\bm{\Xi}_k$ from the priors. 
	
	\item[(b)] Sample cluster weights from $\bm{\eta}_K
 \sim \mathcal{D}(\frac{\alphaM}{K} + |T_1|, \ldots, \frac{\alphaM}{K} + |T_K|)$.

\end{algorithmic}
\end{algorithm}

Note, that in Block $4$, step (a) of Algorithm \ref{Alg1}, $\bm{\Xi}_k$ and $\bm{\theta}_k = \{\theta_{hk}\}$ for the added empty clusters are sampled using, respectively,  $\bm{b}_\xi = (b_{\xi 1}, \ldots, b_{\xi p})$, $b_\theta$ and $b_0$ learned in Block $2$, step (b) of Algorithm \ref{Alg1} from the $K_+$ filled components. This is a specific feature of the telescoping sampler for MFMs developed in \cite{TS2020}, which ensures that the parameters of the filled components inform the parameters of the empty components. 

A separate factor-analytical procedure needs to be run in Block $2$, step (a) for each of the filled clusters $1, \ldots, K_+$ (see Algorithm \ref{Block2a}). The first three steps are standard Gibbs sampler steps for factor models, with the first step used for updating factors $\bm{f}^k_t$ for all observations $t \in T_k$ assigned to cluster $k$. In the following two steps, factor loadings $\bm{\lambda}_{ik}$ in the $i$th row of $\bm{\Lambda}_k$ and idiosyncratic variances $\xi^2_{ik}$ are updated for $i = 1, \ldots, p$. Since classification in Block $1$, step (a) of Algorithm \ref{Alg1} is carried out m.w.r.t. the factors $\bm{f}^k_t$ , it is important to update factors in the first step, so that the factors derived from the observations assigned to the corresponding clusters were used in the subsequent steps updating factor loadings and idiosyncratic variances. In step $4$ of Algorithm \ref{Block2a}, cluster-specific means $\bm{\mu}_k$ are sampled based on the observations assigned to cluster $k$ and the updated parameters of the cluster-specific factor models.

The remaining steps deal with the classification of the columns of the cluster-specific factor loading matrices $\{\bm{\lambda}_{hk}\}, h=1, \ldots, H$ into "active" (assigned to the slab) and "inactive" (assigned to the spike). As already described in Section~\ref{ESP}, this is done by introducing a latent binary indicator $I_{hk}$ for each column $h = 1, \ldots, H$ of each matrix $\bm{\Lambda}_k$, which takes the value of $0$ if the corresponding column $\bm{\lambda}_{hk}$ is assigned to the spike and of $1$ if the corresponding column is assigned to the slab. The classification itself is performed in step $5$, where the values $0$ or $1$ are assigned to $I_{hk}$ according to the marginal probabilities of $\bm{\lambda}_{hk}$ arising from either the spike or the slab distribution.
In step $6$, slab probabilities $\tau_{hk}$ are updated based on the binary indicators $I_{hk}$. Finally, in step $7$, the cluster-specific factor loading variances $\theta_{hk}$ are sampled separately for the columns assigned to the spike and for those assigned to the slab.

Algorithm \ref{Block2b} describes the procedure in Block $2$, step (b) of Algorithm \ref{Alg1}, where the hyperparameteres of the factor-analytical models are updated based on the information derived from the $K_+$ filled clusters in Block $2$, step (a).

\begin{algorithm}[t!]
    \caption{Details of the step (a) in Block 2 of the Algorithm \ref{Alg1}} \label{Block2a}
    \small
 \begin{algorithmic}
 \FOR{$(k \; \text{in} \; 1:K_+)$}
     \begin{enumerate}
      \item Sample $\bm{f}^k_t$ for $t: t \in T_k$ from
\begin{align*}
 \bm{f}^k_t \, | \, - \sim N_H \left((\bm{\Phi}_H + \bm{\Lambda}^T_k\bm{\Xi}_k^{-1}\bm{\Lambda}_k)^{-1}\bm{\Lambda}^T_k\bm{\Xi}_k^{-1}(\bm{y}_t-\bm{\mu}_k), (\bm{\Phi}_H + \bm{\Lambda}^T_k\bm{\Xi}_k^{-1}\bm{\Lambda}_k)^{-1} \right),
\end{align*}
where $\bm{\Xi}_k = \diag(\xi_{1k}^{2}, \ldots, \xi_{pk}^{2})$ and $\bm{\Phi}_H  = \bm{I}_H$.
   	 \item Sample the $i$th row $\bm{\lambda}_{ik}$ of the $k$th cluster loading matrix for $i$ in $(1, \ldots, p)$ from
\begin{align*}
 \bm{\lambda}_{ik}^\top 
 |- \sim N_H \left((\bm{\Psi}_k^{-1} + \xi_{ik}^{-2}
 \bm{F}_k \bm{F}_k^{T})^{-1}\bm{F}_k\xi_{ik}^{-2}(\bm{y}_i-\mu_{ik})^{T}, (\bm{\Psi}_k^{-1} + \xi_{ik}^{-2}\bm{F}_k \bm{F}_k^{T})^{-1} \right),
\end{align*}
where $\bm{\Psi}_k = \diag(\theta_{1k}, \ldots, \theta_{Hk})$, 
$\bm{F}_k = \{ \bm{f}^k_t: t \in T_k\}$
is a matrix of factors of the cluster $k$,  and $\bm{y}_i$ is a vector of observations of the variable $i$, for which $S_t=k$.

	 \item Sample $\xi_{ik}^{-2}$ for $i$ in $(1, \ldots, p)$ from
\begin{align*}
\xi_{ik}^{-2}|- \sim \mathcal{G} \left(a_{\xi}+\frac{|T_k|
}{2},b_{\xi i}+\frac{1}{2}\sum_{t: t \in T_k}(y_{it}-\mu_{ik}-\bm\lambda _{ik}
\bm{f}_t^k)^2\right).
\end{align*}

	 \item Update the cluster-specific mean $\bm{\mu}_k$ from $\bm{\mu}_k\,|\, - \sim N_p(\bm{b}_k, \bm{B}_k)$, where
\begin{align*}
\bm{b}_k = \bm{B}_k \left(\bm{B}^{-1}_0\bm{b}_0 + \bm{\Xi}_k^{-1} \sum_{t: t \in T_k} (\bm{y}_t - \bm{\Lambda}_k \bm{f}_t) \right), \qquad
\bm{B}_k = (\bm{B}^{-1}_0 + |T_k|
\, \bm{\Xi}_k^{-1})^{-1}.
\end{align*}

	 \item Sample the binary indicators $I_{hk}$ for each column $\{ \bm{\lambda}_{hk} \}$, $h = 1, \ldots, H$ of the loading matrix as
 \begin{eqnarray*} 
    P(I_{hk}=0|\bm{\lambda}_{hk}, b_0, \alphaB , H) &\propto &  \frac{H}{\alphaB
    +H}  t_{2a_0} (\bm{\lambda}_{hk};0, (b_0/a_0) \bm{I}_p)\\ 
   P(I_{hk}=1|\bm{\lambda}_{hk}, b_\theta, \alphaB
   , H ) &\propto & \frac{\alphaB}{\alphaB+H} t_{2a_\theta} \left(\bm{\lambda}_{hk};0,(b_\theta/a_\theta) \bm{I}_p \right).
 \end{eqnarray*}
 
 	 	\item  Sample the (unordered) slab probabilities $\tau_{hk}$ for $h$ in $(1, \ldots, H)$:
\begin{align*} 
\tau_{hk} | I_{hk} \sim \mathcal{B} \left(\frac{\alphaB}{H} + I_{hk} , 2 - I_{hk} \right).
\end{align*}

	 \item Given $I_{hk}$ and the $h$th column $\bm{\lambda}_{hk}$ of the loading matrix, for each $h$ in $(1, \ldots, H)$, sample $\theta_{hk}|I_{hk}, \bm{\lambda}_{hk}$ depending on $I_{hk}$:
\begin{align*}
\theta_{hk}|I_{hk}=0, \bm{\lambda}_{hk} \sim  \mathcal{G}^{-1}	\left(a_0 + \frac{1}{2}p, b_0 + \frac{1}{2}\sum_{i=1}^{p}\lambda_{ihk}^{2}\right), \\
\theta_{hk}|I_{hk}=1, \bm{\lambda}_{hk} \sim \mathcal{G}^{-1}\left(a_\theta + \frac{1}{2}p, b_\theta + \frac{1}{2}\sum_{i=1}^{p}\lambda_{ihk}^{2}\right). \nonumber
\end{align*} 
\end{enumerate}
\ENDFOR
\end{algorithmic}
\end{algorithm}
\clearpage

\begin{algorithm}[t!]
    \caption{Details of the step (b) in Block 2 of the Algorithm \ref{Alg1}} \label{Block2b}
    \small
 \begin{algorithmic}[1]

\STATE Sample $b_{\xi i}$ for $i = 1, \ldots, p$ from
\begin{align*}
b_{\xi i} \, | \, - \sim \mathcal{G} \left( a_g + K_+ a_{\xi}, b_{gi} + \sum_{k=1}^{K_+} \frac{1}{\xi^2_{ik}} \right).
\end{align*}

\STATE Calculate the effective number of \lq\lq active\rq\rq\ columns  $H_k$ in cluster $k$ as $H_k
= \sum_{h=1} ^H I_{hk}$. Define $H^{++}=\sum _{k=1}^{K_+} H_k$ as the total number of \lq\lq active\rq\rq\ columns in all filled clusters and $H^\infty=H K_+ -H^{++}$ as the total number of \lq\lq inactive\rq\rq\ columns in all filled clusters.

\STATE Sample $b_0$ from
\begin{align*}
b_0| - \sim \mathcal{G}\left(a_1 + H^\infty a_0, b_1 + \sum _{k=1}^{K_+} \sum_{h: I_{hk}=0} \frac{1}{\theta_{hk}}\right).
\end{align*}
  
\STATE Sample $b_\theta$ from
\begin{align*}
b_\theta| - \sim \mathcal{G}\left(a_2 + H^{++} a_\theta, b_2 + \sum _{k=1}^{K_+} \sum_{h: I_{hk}=1} \frac{1}{\theta_{hk}}\right).
\end{align*}

\STATE  Use a random walk MH step to sample $\alphaB \, | \, H_1, \ldots, H_{K_+}, H
$ from
\begin{align*}
p(\alphaB \, | \, H_1, \ldots, H_{K_+}, H) \propto \left( \frac{\alphaB}{\alphaB+ H} \right)^{H^{++}} \left( \frac{H}{\alphaB + H} \right)^{H^\infty} p(\alphaB).
\end{align*}

\end{algorithmic}
\end{algorithm}

Note that in step~$5$ of Algorithm~\ref{Block2b} we marginalised out $\tau_{hk}$ and sampled $\alphaB$ directly from the information from the classifications of the columns of factor loading matrices into \lq\lq active\rq\rq\ and \lq\lq inactive\rq\rq, namely on the number of active factors in each filled cluster $H_1, \ldots, H_{K_+}$ (see \cite{SFS2022} for the single factor model solution). 
This is done via a random walk Metropolis-Hastings step with proposal $\log \alphaB^{new} \sim \mathcal{N}(\log \alphaB, s^2_\alpha)$. As the acceptance rate depends on the dimension of the data set $p$ through $H$, we made the step size $s_\alpha$ dependent on $H$ exponentially $s_\alpha = 1 + \alpha_1 (1-\alpha_2)^H$, thus making sure that the step size is getting smaller as $p$ (and hence $H$) increases. In our empirical settings we used $\alpha_1=2$ and $\alpha_2=0.11$.

Alternatively, it is also possible to sample $\alphaB$ in a Gibbs sampling step conditioning on $\tau_{hk}$, $H$ and $K_+$ from
\begin{align*} \alphaB \, | \, \tau_{hk}, H, K_+ \sim \mathcal{G} \left(a_\alpha + H K_+ , b_\alpha - \frac{1}{H}\sum_{k=1}^{K_+} \sum_{h=1}^{H} \log \, \tau_{hk} \right).
\end{align*}
However, we found that sampling $\alphaB$ conditional on $\tau_{hk}$ in some cases leads to $\alphaB$ being stuck at relatively high values and results in an overestimation of $H_1, \ldots, H_{K_+}$, while
marginalising out $\tau_{hk}$ leads to a more stable performance of the algorithm.

\subsection{Initialisations and starting values} \label{Init:start}

It is often the case when constructing an MCMC algorithm involving mixture and factor models that starting values are influential in defining the path of the chain. Hence, in order to minimise the probability of the chain being stuck in areas with low posterior probability, initialisations of the model parameters should be chosen carefully. Here we discuss the initialisations and starting values for our model in more details.

The initial splitting of the data into the starting number of clusters $K_0$ is done via k-means clustering
(using R-package \textit{mclust}) to achieve reasonably balanced initial cluster sizes, as using hierarchical clustering to initialise cluster labels often gives heavily imbalanced starting values. $K_0$ is chosen conservatively and should be clearly overfitting. We follow the suggestion in \cite{TS2020} and take it approximately two or three times the expected number of clusters in the data set. Cluster means are initialised as k-means cluster centres and the initial values of cluster weights $\eta_k$, $k=1, \ldots, K$ are sampled from the symmetric Dirichlet distribution $Dir_K(\frac{\alpha}{K})$ with the concentration parameter $\alpha=1$. The Dirichlet concentration parameter $\alphaM$ is initialised as the mean of its prior distribution $\mathcal{F}(\nu_l, \nu_r)$.

The strength parameter of the prior for slab probabilities $\alphaB$ is initiated at the mean of its prior distribution $\mathcal{G}(a_\alpha, b_\alpha)$. The initial allocation of the columns of the cluster-specific factor loading matrices $\bm{\Lambda}_k$ to spike and slab are done according to the slab probabilities $\tau_{hk}$ initiated from the prior. The spike and slab variances $\theta_{hk}$ are initiated as the means of their respective prior distributions.

Special attention should be given to the initialisation of the cluster covariance matrices $\bm{\Omega}_k$. In a classical clustering model they would be given an inverse Wishart prior with some carefully tuned hyperparameters and initialised from this prior. However, in the $(MF)^2 A$ model, the cluster covariance matrices are defined as $\bm{\Omega}_k = \bm{\Lambda}_k \bm{\Lambda}_k^T + \bm{\Xi}_k$ and thus have a far more flexible structure than the inverse Wishart prior. To ensure a proper functioning of the algorithm at the beginning of the chain it is important for the cluster covariance matrices $\bm{\Omega}_k$ to be closely linked to the data.  If both $\bm{\Lambda}_k$ and $\bm{\Xi}_k$ are initiated from the priors, this would very likely take some iterations to achieve and might lead to the chain being stuck in a region of parameter space with low likelihood. As a solution to this problem, we suggest initiating $\bm{\Omega}_k$ for all $k = 1, \ldots, K$ from the estimator suggested in \cite{SFSLopes2018} for the sample precision matrix in the context of sparse Bayesian factor models. This estimator combines the sample information with an inverted Wishart prior $\bm{\Omega}_k \sim \mathcal{IW}_p(v_0, (v_0 \bm{S}_0)^{-1})$.
Provided that the data are standardized, this yields following estimator:  
\begin{align} \label{Omega_est}
\widehat{\bm{\Omega}_k} = (v_0 + T/2)^{-1}(v_0 \bm{S}_0 + 0.5 \sum_{t=1}^T\bm{y}_t\bm{y}_t^T).
\end{align}
Based on hyperparameters $v_0=3$ and $\bm{S}_0 = \bm{I}_p$, the estimator $\widehat{\bm{\Omega}_k}$ is used to initiate the cluster covariance matrix $\bm{\Omega}_k$ for each of the $K$ clusters, thus they all are the same at the first iteration of the MCMC sampler. For unstandardised (but demeaned) data, this estimator can be viewed as an estimator for the sample correlation matrix. In this case the estimator should be appropriately scaled using the diagonal entries of the sample covariance matrix (\cite{Wang2015}). 

\subsection{Post-processing and stratification} \label{Label:switch}

Before conducting any inference, the model output should undergo a post-processing treatment to ensure the correct grouping of the model components into clusters and the correct representation of the parameters of the cluster-specific factor models.

One of the properties of finite mixtures is their invariance to relabelling of the components of the mixture, a  phenomenon first investigated in \cite{RW1984}. This results in a situation that for a mixture distribution with $K$ components there exist up to $K!$ different ways of arranging the components. Therefore, identification of the clusters and cluster-specific parameters requires handling the label switching problem as a post-processing step before conducting any inference on the cluster-specific parameters. Following \cite{SFS2011}, we work with the point process representation of the MCMC draws, choosing only the $\tilde M$ draws with the number of active components equal to the mode $\hat{K}_{+}$ of $K_+$ and clustering them together. To include information from the cluster-specific factor models, we replace the $k$-means clustering with clustering around $(\bm{\mu}_k ^\top \, \log|\bm{\Omega}_k| \, \log(tr(\bm{\Omega}_k)) \, \log \left(\dfrac{v_k^{\max}} {v_k^{\min}} \right))^\top$, where $v_k^{\max}$ and $v_k^{\min}$ are the biggest and the smallest eigenvalues of $\bm{\Omega}_k$. 

This produces a classification index $J^{(m)}_k \in \{1, \ldots,\hat{K}_{+} \}$ for each of the maintained draws $m = 1, \ldots, \tilde M$. If $\rho_m = (J^{(m)}_1, \ldots, J^{(m)}_{\hat{K}_{+}})$ is a permutation of $\{ 1, \ldots, K_+\}$, a unique labelling is achieved and the cluster-specific model parameters and the latent cluster allocation indicators $S_t ^{(m)}, t= 1, \ldots, T$ are reordered through $\rho_m$. The draws corresponding to $\rho_m$s which are not a permutation of $\{ 1, \ldots, \hat{K}_{+}\}$ are then removed.

Once the cluster assignment is completed, we compute the inferred number of factors $\hat{H}_k$ in each of the clusters as the mode of the number of active factors in each cluster over the draws maintained after the clustering assignment and remove those draws in which the number of active factors in the corresponding clusters is not equal to $\hat{H}_k$. 
We denote the number of remaining posterior draws by $M$.
 Thus, at this stage, our cluster-specific factor loading matrices $\bm{\Lambda}_k$ have $\hat{H}_k$ active columns (with the respective indicator $I_{hk}=1$) and $H-\hat{H}_k$ inactive columns (with the respective indicator $I_{hk}=0$). Correspondingly, cluster-specific matrices of factors $\bm{F}_k$ have $\hat{H}_k$ number of active rows and $H-\hat{H}_k$ number of inactive rows. We keep only the active columns of $\bm{\Lambda}_k$ and rows of $\bm{F}_k$ (which correspond to the binary indicator $I_{hk}=1$). The resulting cluster specific factor loading matrices $\bm{\Lambda}_k = (\bm{\lambda}_1, \ldots, \bm{\lambda}_{\hat{H}_k})$ with $\hat{H}_k$ columns are then used for the calculation of cluster-specific covariance matrices in Section \ref{Simulation}.

\section{Simulation studies} \label{Simulation}

The performance of the dynamic $(MF)^2 A$ model is first demonstrated on simulation studies. We use several different settings to assess the model's ability to correctly infer the cluster and factor dimensionality of the data sets. In Section \ref{study1} we demonstrate the performance of the model for a range of various $p$ and $T$ settings on data sets with balanced cluster sizes and a common number of factors. The simulation study in Section \ref{study2} is more challenging with a larger number of clusters, some of which are small, and a varying number of cluster-specific factors. In Section \ref{IMIFA:sim} we compare the performance of the dynamic $(MF)^2 A$ model and the IMIFA model on the data sets used in Sections \ref{study1} and \ref{study2}.

Unless otherwise stated, data are standardised, which means mean-centred and unit-scaled. The hyperparameter specifications are reported in Table~\ref{tab:hyperpar}. The maximum number of factors $H$ is equal to the smallest integer which satisfies the variance identification condition of \cite{Anderson1956} $H \leq \frac{p-1}{2}$. The only exception is the smallest setting of $(p, T)$ in the Simulation Study 1, where $p=10$, in which case $H=p$ was used. 
Unless otherwise specified, the sampler is run for $50,000$ iterations, with $20$\% of them discarded as  burn-in.  To test the robustness of the model's performance, each simulation setting was replicated five times and each time the data set was newly generated. 

The clustering performance is assessed using the adjusted Rand index (ARI; \cite{ARI1985}) and the misclassification rate is estimated as the percentage of mislabelled observations compared to the true cluster labels used to simulate the data.
 To assess the accuracy of the model in estimating the true cluster-specific covariance matrices $\bm{\Omega}_k^0 = \bm{\Lambda}_k^0 (\bm{\Lambda}_k^{0} ) ^\top + \bm{\Xi}_k^0$ of the data via the estimated cluster-specific covariance matrices 
 $$\widehat{\bm{\Omega}}_k = \frac{1}{M} \sum_{m=1}^M \bm{\Omega}_k ^{(m)},
 \quad \bm{\Omega}_k ^{(m)}=
\bm{\Lambda}_k^{(m)} (\bm{\Lambda}_k^{(m)})^\top + \bm{\Xi}_k ^{(m)},$$
where $\bm{\Lambda}_k^{(m)}$ and $\bm{\Xi}_k^{(m)}$ are the $m$-th among $M$ posterior draws left after applying the post-processing procedure described in Section \ref{Label:switch}, we compute for each simulation in each of the scenarios a Monte-Carlo estimate of the mean squared error (MSE) defined by
\begin{align*}
MSE_{\Omega_k} = \sum_{i=1}^p \sum_{l=i}^p \mathop{\mathbb{E}} ((\widehat{\Omega}_{k, il} - \Omega^0_{k, il})^2 \, | \, \bm{y})/(p(p + 1)/2).
\end{align*}

Following \cite{Murphy2019}, the data are standardised before feeding them into the model. More specifically, the data are transformed as $\tilde{\bm{y}}_t= \bm{S}^{-1} (\bm{y}_t - \bm{m})$, where $\bm{m}$ denotes the vector of means of $\bm{y}$ and the scale matrix $\bm{S}=\sqrt{diag(S_{y,1} \ldots S_{y,p} )}$ is defined from the empirical variances $S_{y,i}$ of the data $y_{it}$ over $t=1, \ldots,T$. Given the true cluster-specific factor loading and 
covariance matrices $\bm{\Lambda}_k^0$ and $\bm{\Omega}_k^0$ 
of the original data, 
the corresponding matrices 
then take the form  $\bm{S}^{-1} \bm{\Lambda}_k^0$
and $\bm{S}^{-1} \bm{\Omega}^0_k \bm{S}^{-1}$ for the transformed data.

\subsection{Simulation study 1} \label{study1}
	
The aim of this simulation study is to evaluate the performance of our model on data sets of various sizes, i.e. with various settings of $p$ and $T$,  with the clusters approximately equally sized but not very well separated from each other. Three different settings of $(p, T)$ were considered to test the performance of the model on small, middle sized and relatively large data sets, and also to evaluate the results against an increasing number of observations, namely $(10, 100)$, $(30, 200)$ and $(50, 500)$.  Note, that for a reliable performance of the model, the number of observations should be reasonably bigger than the number of variables.

\begin{figure}[t!]
  \centering
  \includegraphics[width=0.75\textwidth, height=0.75\textwidth]{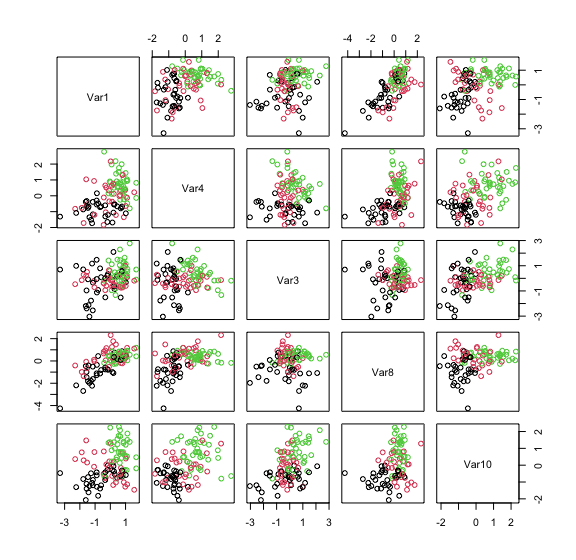}
  \vspace{-10mm}
  \caption{Pairwise scatterplots of $5$ randomly chosen variables from one of the replicate data sets in Simulation Study 1 with $p=10$ and $T=100$ demonstrating the overlap between $3$ clusters.} 
  \label{fig:sim1}
\end{figure}

The data are simulated with $K_+ = 3$ clusters and $H_k=4$ factors in each cluster, and with the cluster weights $\bm{\eta}_K = (1/3, 1/3, 1/3)$. Other model parameters are simulated as $\bm{f}^k_t \sim N_{H_k} (\bm{0}, \bm{I}_{H_k})$, $\xi^2_{ik} \sim \mathcal{G}^{-1}(2, 1)$, and $\lambda_{ik} \sim N_{H_k} (\bm{0}, \bm{I}_{H_k})$ for all $k$. To ensure that clusters are overlapping, the means are generated in the following way, similar to \cite{Murphy2019}: $\bm{\mu}_k \sim N_p((2k-K_+-1)\bm{1}, \bm{I}_p)$. The data $\bm{y}$ are simulated according to the conditional mixture model
\begin{align*}
p(\bm{y}_t \, | \{ \, \bm{\mu}_k, \bm{\Lambda}_k, \bm{f}^k_t, \bm{\Xi}_k\}, \bm{\eta}_K ) = \sum_{k=1}^{K_+} \eta_k N_p\left(\bm{y}_t; \bm{\mu}_k + \bm{\Lambda}_k \bm{f}^k_t, \bm{\Xi}_k\right).
\end{align*}
As the cluster-specific $\bm{\Lambda}_k$ and $\bm{\Xi}_k$ parameters could induce some degree of separation between clusters, pairwise scatterplots from a randomly chosen raw data set is shown in Figure \ref{fig:sim1} to demonstrate the extent of overlap amongst clusters. For the sake of clear visibility, five randomly chosen variables from a data set with $p=10$ variables and $T=100$ observations are depicted.

\renewcommand{\arraystretch}{1.5}
\begin{table}[t!]
  \begin{center}
    \resizebox{\columnwidth}{!}{
    \begin{tabular}{c|c|c|c|c|c|c|c|c|c|c} 
    \hline
      Dimension & $\hat{K_+}$ & $\hat{K}$ & $\hat{H}_1$ & $\hat{H}_2$ & $\hat{H}_3$ &  ARI & Error (\%) & $MSE_{\Omega_1}$ & $MSE_{\Omega_2}$ & $MSE_{\Omega_3}$ \\
      \hline 
      $(10, 100)$ & 3 [3,3] & 3 [3,9] & 4 [4,5] & 3 [3,5] & 4 [4,4] & 1 & 0 & 0.076 & 0.031 & 0.018 \\
       & 3 [3,3] & 3 [3,9] & 4 [4,4] & 4 [4,4] & 4 [3,4] & 1 & 0 & 0.032 & 0.023 & 0.063 \\
       & 3 [3,3] & 3 [3,9] & 3 [3,4] & 4 [4,5] & 4 [4,5] & 0.97 & 1 & 0.016 & 0.023 & 0.047 \\
       & 3 [3,3] & 3 [3,9] & 4 [4,6] & 4 [3,5] & 4 [3,5] & 1 & 0 & 0.059 & 0.010 & 0.009 \\
       & 3 [3,3] & 3 [3,9] & 4 [4,5] & 4 [4,5] & 4 [4,6] & 1 & 0 & 0.024 & 0.046 & 0.042 \\
      \hline 
      $(30, 200)$ & 3 [3,3] & 3 [3,8] & 4 [4,4] & 4 [4,4] & 4 [4,4] & 1 & 0 & 0.009 & 0.012 & 0.010 \\
      & 3 [3,3] & 3 [3,8] & 4 [4,4] & 4 [4,4] & 4 [4,4] & 1 & 0 & 0.018 & 0.019 & 0.015 \\
       & 3 [3,3] & 3 [3,8] & 4 [4,4] & 4 [4,4] & 4 [4,4] & 1 & 0 & 0.008 & 0.012 & 0.011 \\
       & 3 [3,3] & 3 [3,8] & 4 [4,4] & 4 [4,4] & 4 [4,4] & 1 & 0 & 0.012 & 0.009 & 0.010 \\
       & 3 [3,3] & 3 [3,8] & 4 [4,4] & 4 [4,4] & 4 [4,4] & 1 & 0 & 0.015 & 0.012 & 0.016 \\
      \hline
      $(50, 500)$ & 3 [3,3] & 3 [3,7] & 4 [4,4] & 4 [4,4] & 4 [4,4] & 1 & 0 & 0.003 & 0.006 & 0.004 \\
       & 3 [3,3] & 3 [3,8] & 4 [4,4] & 4 [4,4] & 4 [4,4] & 1 & 0 & 0.005 & 0.004 & 0.008 \\
       & 3 [3,3] & 3 [3,7] & 4 [4,4] & 4 [4,4] & 4 [4,4] & 1 & 0 & 0.006 & 0.008 & 0.005 \\
       & 3 [3,3] & 3 [3,8] & 6 [6,6] & 5 [5,5] & 4 [4,4] & 1 & 0 & 0.005 & 0.005 & 0.005 \\
       & 3 [3,3] & 3 [3,7] & 4 [4,4] & 4 [4,4] & 4 [4,4] & 1 & 0 & 0.004 & 0.005 & 0.004 \\
      \hline
    \end{tabular} }
  \end{center}
  \vspace{-5mm}
  \caption{Simulation results for the $(MF)^2A$ model under different dimensionality settings. The modal estimates of $K$ and $H_k$ are reported, with 95\% credible intervals given in brackets. Clustering performance is assessed via the ARI and the average percentage error rate against the known cluster labels using the R-package \textit{mclust}. The true numbers are $K_+=3$ and $H_k=4$.}
  	\label{tab:sim1}
\end{table}

The results provided in Table~\ref{tab:sim1} demonstrate that the dynamic $(MF)^2 A$ model performs generally well for all three settings of $(p,T)$, exhibiting the capability to uncover the true structure of the simulated data in most cases. The partition has been identified correctly in all cases except one case in the smallest $(p, T)$ setting, where one observation was misclassified and led to the ARI of $0.97$. The number of cluster-specific factors was occasionally slightly underestimated in the $(10,100)$ setting, however in higher $(p, T)$ settings the correct number of cluster-specific factors was identified correctly for almost all replicate data sets. In general, the model exhibited a stable performance both on small and relatively large data sets.
	
	\subsection{Simulation study 2} \label{study2}
	
\begin{figure}[t!]
  \centering
  \includegraphics[width=0.80\textwidth, height=0.80\textwidth]{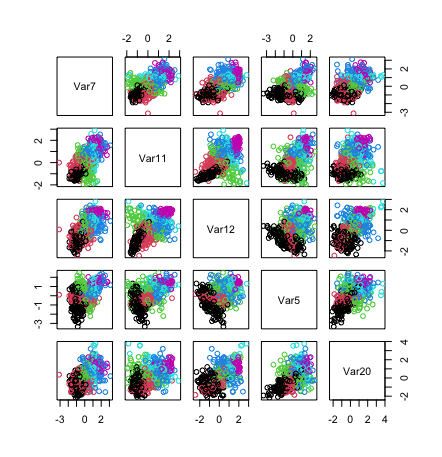}
  \vspace{-10mm}
  \caption{Pairwise scatterplots of 5 randomly chosen variables from one of the replicate data sets in Simulation Study 2 with $p=20$ and $T=700$ demonstrating the overlap between 6 clusters.} 
  \label{fig:sim2}
\end{figure}

The design of the simulation study presented in this section is more challenging for the algorithm as the clusters are of different sizes and the number of factors varies between clusters. The data is generated with $p=20$ and  $T=700$, and is allocated into $K_+ = 6$ clusters with varying numbers of cluster-specific factors. The clusters are given weights $\bm{\eta}_6 = (0.25, 0.25, 0.2, 0.15, 0.1, 0.5)$, thus including rather small clusters (a setting which often appears in Bayesian nonparametric models). The number of factors $H_1, \ldots, H_{K_+}$ are drawn randomly from $1, \ldots, 5$,
with the upper limit being smaller than $\frac{p-1}{2}$ and thus satisfying the variance identification constraint of \cite{Anderson1956}. Otherwise, the same parameter settings as in the Simulation Study 1 in Section~\ref{study1} were used to generate the data. Figure \ref{fig:sim2} illustrates the extent of intermixing between the clusters by showing pairwise scatterplots for five randomly chosen variables for the first replicate data set.

The sampler was run for $25,000$ iterations, with $20\%$ of them discarded as burn-in. The parameters, namely the number of clusters $K_+$, the partition $\mathcal{C}$ and the cluster-specific number of factors $\bm{H} = (H_1, \ldots, H_{K_+})$, of the five simulated data sets are summarised in Table \ref{tab:sim2}. The estimated parameters as well as the ARI and the clustering error are presented in Table~\ref{tab:mod2}.
 
\renewcommand{\arraystretch}{1.3}
\begin{table}[t!]
  \begin{center}
    \small
    \begin{tabular}{c|c|c|c} 
    \hline
      & $K_+$ & $\mathcal{C}$ & $\bm{H}$  \\
      \hhline{=|=|=|=}
      1st replicate & 6 & (183, 160, 158, 94, 71, 34) & (3, 3, 4, 1, 4, 2) \\
       \hline
       2nd replicate & 6 & (180, 195, 131, 90, 68, 36) & (2, 4, 2, 4, 4, 1) \\
       \hline
       3d replicate & 6 & (197, 155, 131, 104, 71, 42) & (4, 2, 3, 5, 4, 5) \\
       \hline
       4th replicate & 6 & (177, 169, 136, 108, 72, 38) & (1, 3, 1, 2, 3, 3) \\
       \hline
       5th replicate & 6 & (182, 174, 144, 102, 60, 38) & (2, 3, 5, 1, 4, 5) \\
       \hline
     \end{tabular}
\end{center}
  \vspace{-5mm}
  \caption{Parameters of five simulated data sets in the situation with unbalanced cluster sizes and different number of cluster-specific factors.}
  	\label{tab:sim2}
\end{table}

\renewcommand{\arraystretch}{1.3}
\begin{table}[t!]
  \begin{center}
    \small
    \begin{tabular}{c|c|c|c|c|c} 
    \hline
      & $\hat{K}_+$ & $\hat{\mathcal{C}}$ & $\hat{\bm{H}}$ & ARI & Error (\%) \\
      \hhline{=|=|=|=|=|=}
      1st replicate & 6 & (183, 160, 158, 94, 71, 34) & (3, 3, 5, 1, 4, 2) & 1 & 0 \\
       \hline
       2nd replicate & 6 & (180, 195, 131, 90, 68, 36) & (2, 4, 2, 4, 4, 1) & 1 & 0 \\
       \hline
       3d replicate & 5 & (197, 155, 235, 71, 42) & (4, 2, 9, 4, 5) & 0.85 & 14.8 \\
       \hline
       4th replicate & 6 & (177, 169, 136, 108, 72, 38) & (1, 3, 1, 3, 3, 4) & 1 & 0 \\
       \hline
       5th replicate & 6 & (182, 174, 144, 102, 60, 38) & (2, 3, 5, 1, 5, 6) & 1 & 0 \\
       \hline
     \end{tabular}
\end{center}
  \vspace{-5mm}
  \caption{Estimation results for the $(MF)^2A$ model in the situation with unbalanced cluster sizes and different number of cluster-specific factors.}
  	\label{tab:mod2}
\end{table}

In four out of five cases, the model correctly identified the partition and cluster assignments. However, in the case of the third replicate data set, the $3$d and the $4$th clusters were joined together. Regarding the inference on the number of cluster-specific factors, in most cases the number of cluster-specific factors was determined correctly with only occasional slight overestimation, mostly in the case of smaller cluster sizes. Summing up, the results show that the model performs reliably well in situations with unbalanced cluster sizes, but sometimes overestimates the number of factors in small clusters (when the gap between the number of variables $p$ and the number of observations $|T_k|$ is relatively small).

\subsection{Simulation study 3: comparison with IMIFA model} \label{IMIFA:sim}

To compare our dynamic $(MF)^2A$ model with its closest counterpart, we run the IMIFA model of \cite{Murphy2019}, fit via the R-package \textit{IMIFA}, on exactly the same simulation data sets as in the previous two sections. To keep the settings of the two models as close as possible, we set the PYP parameters of the IMIFA model, namely $\sigma$ and $d$, to be learned from data and run the model for exactly the same number of iterations. The MGP parameters are left as default, i.e. $a_1 = 2.1$ and $a_2 = 3.1$.  Table~\ref{Sim1:IMIFA} presents the results of running the IMIFA model on the data sets with $K_+ = 3$ clusters, a common number of factors $H_k = 4$ in all clusters, and various settings of $(p, T)$, exactly as in Simulation Study~1 in Section~\ref{study1}.

\renewcommand{\arraystretch}{1.3}
\begin{table}[t!]
  \begin{center}
    \small
    \begin{tabular}{c|c|c|c|c|c|c} 
    \hline
      Dimension & $\hat{K}_+$ & $\hat{H}_1$ & $\hat{H}_2$ & $\hat{H}_3$ &  ARI & Error (\%) \\
      \hline 
      $(10, 100)$ & 3 [3,3] & 5 [3,7] & 5 [4,7] & 4 [3,6] &  1 & 0 \\
       & 3 [3,3] & 5 [3,6] & 4 [3,6] & 4 [2,6] & 1 & 0 \\
       & 3 [3,3] & 4 [3,6] & 4 [3,6] & 5 [4,7] & 0.97 & 1 \\
       & 3 [3,3] & 5 [4,7] & 4 [2,5] & 4 [2,5] & 1 & 0 \\
       & 3 [3,3] & 5 [3,7] & 5 [3,7] & 4 [3,7] & 1 & 0 \\
      \hline 
      $(30, 200)$ & 3 [3,3] & 5 [4,6] & 5 [4,6] & 5 [4,6] & 1 & 0 \\
      & 3 [3,3] & 5 [4,6] & 5 [4,6] & 5 [4,7] & 1 & 0 \\
       & 3 [3,3] & 5 [4,6] & 5 [4,6] & 5 [4,6] & 1 & 0 \\
       & 3 [3,3] & 5 [4,6] & 5 [4,6] & 5 [4,6] & 1 & 0 \\
       & 3 [3,3] & 5 [4,6] & 5 [4,6] & 5 [4,6] & 1 & 0 \\
      \hline
      $(50, 500)$ & 3 [3,3] & 5 [4,6] & 5 [4,6] & 5 [4,6] & 1 & 0 \\
       & 3 [3,3] & 5 [4,6] & 5 [4,6] & 5 [4,6] & 1 & 0 \\
       & 3 [3,3] & 5 [4,6] & 5 [4,7] & 5 [4,6] & 1 & 0 \\
       & 3 [3,3] & 5 [4,6] & 5 [4,6] & 5 [4,6] & 1 & 0 \\
       & 3 [3,3] & 5 [4,6] & 5 [4,6] & 5 [4,6] & 1 & 0 \\
      \hline
    \end{tabular}
  \end{center}
  \vspace{-5mm}
  \caption{Simulation results for the IMIFA model under different dimensionality settings. The modal estimates of $K$ and $H_k$ are reported, with 95\% credible intervals given in brackets. Clustering performance is assessed via the ARI and the average percentage error rate against the known cluster labels using the R-package \textit{mclust}.
  The true numbers are $K_+=3$ and $H_1=H_2=H_3=4$.
  }
  	\label{Sim1:IMIFA}
\end{table}

Table~\ref{Sim1:IMIFA} shows that the clustering performance is, in general, quite good, especially for higher values of $(p, T)$. In the factor-analytical part, however, IMIFA shows a clear tendency to some overestimation of the cluster-specific number of factors compared to the results obtained in Table~\ref{tab:sim1} for the dynamic $(MF)^2A$ model, especially on larger data sets. This can be attributed to the intrinsic inefficiency of the MGP prior described in Section~\ref{IMIFA-model}.

\renewcommand{\arraystretch}{1.3}
\begin{table}[t!]
  \begin{center}
    \small
    \begin{tabular}{c|c|c|c|c|c} 
    \hline
      & $\hat{K}_+$ & $\hat{\mathcal{C}}$ & $\hat{\bm{H}}$ & ARI & Error (\%) \\
      \hhline{=|=|=|=|=|=}
      1st replicate & 6 & (183, 160, 158, 94, 71, 34) & (4, 4, 5, 2, 5, 3) & 1 & 0 \\
       \hline
       2nd replicate & 6 & (180, 195, 131, 90, 68, 36) & (3, 5, 3, 5, 5, 2) & 1 & 0 \\
       \hline
       3d replicate & 6 & (197, 155, 131, 104, 71, 42) & (5, 3, 4, 6, 5, 6) & 1 & 0 \\
       \hline
       4th replicate & 6 & (177, 169, 136, 108, 72, 38) & (2, 4, 2, 3, 4, 4) & 1 & 0 \\
       \hline
       5th replicate & 6 & (182, 174, 144, 102, 60, 38) & (3, 4, 6, 2, 5, 6) & 1 & 0 \\
       \hline
     \end{tabular}
\end{center}
  \vspace{-5mm}
  \caption{Estimation results for the IMIFA model in the situation with unbalanced cluster sizes and different number of cluster-specific factors.}
  	\label{Sim2:IMIFA}
\end{table}

The results of running the IMIFA model on the simulation data sets with different number of factors (as in Table~\ref{tab:sim2}) are presented in Table~\ref{Sim2:IMIFA}. They confirm the tendency of the model to overestimate the number of factors, which was evident in the results in Table~\ref{Sim1:IMIFA}. The clustering performance is perfect and smaller clusters do not represent a problem, which can be explained by the general good performance of PYP prior in situations with a large number of smaller clusters.

\section{Applications} \label{Applications}

In this section we evaluate the performance of the dynamic $(MF)^2A$ on several publicly available benchmark data sets, which are often used in the literature to test MFA models. We compare the performance of our model against three other MFA models, the first of which is IMIFA, fit via the R package \textit{IMIFA} (\cite{Murphy2019}), and which is the only one that can be compared with our model in terms of flexibility in a sense that it also provides fully automatic inference on the number of clusters and cluster-specific factors and allows factors to differ between clusters. The other two models, namely overfitting Bayesian mixtures of factor analysers, fit via the R package \textit{fabMix} (\cite{Papastamoulis2020}), and parsimonious Gaussian mixture models, fit via the R package \textit{pgmm} (\cite{MCNM2018}), are less flexible and only allow the same number of factors in each cluster. For the sake of simplicity, we will refer to those models with the names of the R packages which were used to fit them, namely \textit{IMIFA}, \textit{fabMix} and \textit{pgmm}. Unless otherwise specified, the data are standardised (demeaned and unit-scaled) before fitting our Bayesian MFA model and the same values of hyperparameters are used as reported in Table \ref{tab:hyperpar} for the simulated data. Unless otherwise specified, the sampler is run for $50,000$ iterations with $20$\% of them discarded as a burn-in.

\subsection{Benchmark data: Coffee data set} \label{coffee}

The coffee data set, first introduced in \cite{Coffee1973}, is one of the benchmark data sets often used to evaluate the performance of clustering and MFA models (see, e.g. \cite{MCNM2008}, \cite{Papastamoulis2018}) and is available in the \textit{pgmm} R package. The data consists of $T = 43$ coffee samples from 29 countries collected from beans corresponding to the Arabica and Robusta species. For each sample 13 variables are observed: water, pH value, fat, chlorogenic acid, bean weight, free acid, caffeine, neochlorogenic acid, extract yield, mineral content, trigonelline, isochlorogenic acid and total chlorogenic acid. Following \cite{MCNM2008}, we excluded the total chlorogenic acid from the analysed data since it is the sum of the chlorogenic, neochlorogenic and isochlorogenic acids, so in the end there are $p=12$ variables in the data set.

Table \ref{tab:coffee} illustrates the performance of all four models in estimation of the coffee data set. It is natural to assume that different coffee bean species, namely Arabica and Robusta, should correspond to different clusters. The ARI and the misclassification rate are computed on the basis of the known classifications into Arabica and Robusta coffee bean species.

\renewcommand{\arraystretch}{1.3}
\begin{table}[t!] 
  \begin{center}
    \small
    \begin{tabular}{c|c|c|c|c} 
    \hline
      Model & \# clusters & \# factors & ARI & Error (\%) \\
      \hhline{=|=|=|=|=}
      Dynamic $(MF)^2A$ & 2 & 1, 2 & 1 & 0 \\
      \hline
      IMIFA & 2 & 3, 5 & 1 & 0 \\
      \hline
      \textit{pgmm} & 5 & 1 
      & 0.32 & 34.9 \\
      \hline
      \textit{fabMix} & 2 & 1
      & 1 & 0 \\
      \hline
    \end{tabular}
  \end{center}
  \vspace{-5mm}
  \caption{Results of fitting the dynamic $(MF)^2A$ model against a range of benchmark MFA models on the coffee data set. Note that the number of factors is estimated to be same for all clusters in \textit{pgmm} and \textit{fabMix} by model design.}
  	\label{tab:coffee}
\end{table}	

All models except the parsimonious Gaussian mixture model (\textit{pgmm}), were able to identify the correct number of clusters and to uncover the correct partition. Here it must be mentioned that the parsimonious Gaussian mixture model is very sensitive to the specification of the initial range of possible values for the number of clusters and factors which the model explores as well as the choice of either random or $k$-means starting points (this problem has also been mentioned in \cite{Papastamoulis2018}).
 For example, the classification results for the coffee data set with two clusters and the correct partition, which are reported in \cite{MCNM2008}, emerge when the range of possible number of cluster is set between $2$ and $3$ and the range of possible number of factors is set between $1$ and $3$. However, having found that choice rather restrictive, we set a slightly wider range of one to five for both the number of clusters and cluster-specific factors\footnote{We used the same range of possible values for the number of cluster-specific factors for the simulation with overfitting Bayesian mixtures of factor analysers via \textit{fabMix} package.}, thus aiming for somewhat more flexibility. The model chosen on the basis of the integrated complete-data likelihood (ICL) criterion reports $5$ clusters with $1$ factors in each cluster, splitting the "Arabica" beans into three groups and "Robusta" beans into two.

\subsection{Benchmark data: Italian wines} \label{wines}

The Italian wines data set (\cite{Forina1986}), available in the \textit{pgmm} R package, is another benchmark data set employed for assessing the performance of clustering and MFA models (see, e.g. \cite{Papastamoulis2018}, \cite{MCNM2008}). It contains $p=27$ variables measuring chemical and physical properties of $T=178$ wines collected over the period $1970-1979$. The wines originate from the Piemont region of Italy and belong to one of the three types: Barolo, Grignolino and Barbera. We expect the classification algorithm to recognise three clusters which correspond to the three wine types.

\renewcommand{\arraystretch}{1.3}
\begin{table}[t!]
  \begin{center}
    \small
    \begin{tabular}{c|c|c|c|c} 
    \hline
      Model & \# clusters & \# factors & ARI & Error (\%) \\
      \hhline{=|=|=|=|=}
      Dynamic $(MF)^2A$ & 4 & 4, 1, 4, 1 & 0.48 & 33.1 \\
      \hline
      IMIFA & 10 & 3, 6, 4, 5, 3, 2, 3, 5, 3, 4 & 0.72 & 19.1 \\
      \hline
      \textit{pgmm} & 3 & 4 & 0.96 & 1.1 \\
      \hline
      \textit{fabMix} & 5 & 1 & 0.66 & 18.0 \\
      \hline
    \end{tabular}
  \end{center}
  \vspace{-5mm}
  \caption{Results of fitting the dynamic $(MF)^2A$ model against a range of benchmark MFA models on the Italian wines data set. Note that the number of factors is estimated to be same for all clusters in \textit{pgmm} and \textit{fabMix} by model design.}
  	\label{tab:wines}
\end{table}	

\renewcommand{\arraystretch}{1.5}
\begin{table}[t!]
  \begin{center}
    \small
    \begin{subtable}{0.33\linewidth}
    \centering
    \begin{tabularx}{\textwidth}{@{} X *4{c} @{}}
\toprule
  & \multicolumn{4}{c}{Dynamic $(MF)^2A$} \\ \cmidrule(r){2-5} 
Cluster & 1 & 2 & 3 & 4   \\ \midrule
Barolo & 0 & 2 & 57 & 0   \\ 
Grignolino & 0 & 2 & 65 & 4   \\
Barbera & 48 & 0 & 0 & 0   \\ \bottomrule
    \end{tabularx} 
    \end{subtable}
\hfil
\vspace{0.5cm}
    \begin{subtable}{0.58\linewidth}
    \centering
    \begin{tabularx}{\textwidth}{Xcccccccccc} \toprule
  & \multicolumn{10}{c}{IMIFA} \\ \cmidrule(r){2-11} 
Cluster & 1 & 2 & 3 & 4 & 5 & 6 & 7 & 8 & 9 & 10  \\ \midrule
Barolo & 49 & 0 & 0 & 0 & 7 & 0 & 0 & 0 & 3 & 0  \\ 
Grignolino & 0 & 54 & 12 & 0 & 1 & 0 & 0 & 1 & 0 & 3  \\
Barbera & 0 & 0 & 0 & 41 & 0 & 5 & 2 & 0 & 0 & 0  \\ \bottomrule
    \end{tabularx} 
    \end{subtable}  
\hfil
    \begin{subtable}{0.4\linewidth}
    \centering
    \begin{tabularx}{\textwidth}{@{} X *3{c} @{}}
\toprule
   & \multicolumn{3}{c}{\textit{pgmm}} \\ \cmidrule(r){2-4} 
Cluster  & 1 & 2 & 3    \\ \midrule
Barolo & 0 & 59 & 0   \\ 
Grignolino & 1 & 1 & 69  \\
Barbera & 48 & 0 & 0  \\ \bottomrule
    \end{tabularx} 
    \end{subtable}
\hfil
\vspace{0.5cm}
    \begin{subtable}{0.5\linewidth}
    \centering
    \begin{tabularx}{\textwidth}{@{} X *5{c} @{}}
\toprule
  & \multicolumn{5}{c}{\textit{fabMix}} \\ \cmidrule(r){2-6} 
Cluster & 1 & 2 & 3 & 4 & 5  \\ \midrule
Barolo & 54 & 4 & 1 & 0 & 0  \\ 
Grignolino & 9 & 3 & 14 & 1 & 44  \\
Barbera & 0 & 0 & 0 & 48 & 0  \\ \bottomrule
    \end{tabularx} 
    \end{subtable}  
  \end{center}
  \vspace{-5mm}
  \caption{Confusion matrices between the estimated and true cluster assignments of the Italian wines data set. The estimated cluster assignments are provided by $(MF)^2A$, IMIFA, \textit{pgmm} and \textit{fabMix} models.}
  	\label{table:conf:MF2A}
\end{table}

The results of applying the dynamic $(MF)^2A$ model and the three alternative MFA models to the Italian wines data set are presented in Table \ref{tab:wines}. The true cluster assignments were computed on the basis of the known classifications into Barolo, Grignolino and Barbera wine types. The confusion matrices between the estimated and the true cluster assignments are given in Table \ref{table:conf:MF2A}.

The best clustering performance is delivered by the \textit{pgmm} model, which produced an almost perfect classification, while the other three models overestimated the number of clusters. The dynamic $(MF)^2A$ model essentially put most of Barolo and Grignolino wines in one cluster and the Barbera wines into a separate cluster. Similarly, the \textit{fabMix} model put all observations belonging to Barbera wine type to a separate cluster but struggled with Barolo and especially Grignolino wines spreading them across four other clusters. The IMIFA model estimated $10$ clusters, with most of the observations being concentrated in four clusters. Regarding the number of cluster-specific latent factors, both the dynamic $(MF)^2A$ and the \textit{pgmm} models estimated $4$ factors in each (significantly filled) cluster. The estimated number of factors by the IMIFA model in bigger clusters is between $3$ and $6$, while the \textit{fabMix} found only one latent factor.

\begin{table}[t!]
  \begin{center}
    \small
    \begin{tabular}{c|c|c|c|c|c|c} 
    \hline
      Model & \# clusters & \# factors & ARI & Error & ARI & Error \\
       &  &  & (areas) & (areas, \%) & (regions) & (regions, \%) \\
      \hhline{=|=|=|=|=|=|=}
      Dynamic $(MF)^2A$ & 5 & 2, 1, 1, 4, 3 & 0.60 & 32.5 & 0.77 & 26.7 \\
      \hline
      IMIFA & 5 & 2, 3, 3, 6, 3 & 0.90 & 17.1 & 0.54 & 27.3 \\
      \hline
      \textit{pgmm} & 5 & 5 & 0.59 & 33.4 & 0.76 & 27.1 \\
      \hline
      \textit{fabMix} & 5 & 4 & 0.59 & 31.8 & 0.76 & 26.9\\
      \hline
    \end{tabular}
  \end{center}
  \vspace{-5mm}
  \caption{Results of fitting the dynamic $(MF)^2A$ model against a range of benchmark MFA models on the Italian olive oils data set. Note that the number of factors is estimated to be same for all clusters in \textit{pgmm} and \textit{fabMix} by model design.}
  	\label{tab:olive}
\end{table}

\renewcommand{\arraystretch}{1.5}
\begin{table}[t!]
  \begin{center}
    \small
    \begin{subtable}{0.45\linewidth}
    \centering
    \begin{tabularx}{\textwidth}{@{} X *5{c} @{}}
\toprule
  & \multicolumn{5}{c}{Dynamic $(MF)^2A$} \\ \cmidrule(r){2-6} 
Cluster & 1 & 2 & 3 & 4 & 5  \\ \midrule
Northern Italy & 0 & 91 & 0 & 60 & 0 \\ 
Sardinia & 0 & 0 & 0 & 0 & 98 \\ 
Southern Italy & 197 & 0 & 126 & 0 & 0 \\  \bottomrule
    \end{tabularx} 
    \end{subtable}
\hfil
\vspace{0.5cm}
    \begin{subtable}{0.45\linewidth}
    \centering
    \begin{tabularx}{\textwidth}{@{} X *5{c} @{}} \toprule
  & \multicolumn{5}{c}{IMIFA} \\ \cmidrule(r){2-6} 
Cluster & 1 & 2 & 3 & 4 & 5  \\ \midrule
Northern Italy & 48 & 0 & 50 & 0 & 53  \\
Sardinia & 0 & 98 & 0 & 0 & 0  \\ 
Southern Italy & 0 & 0 & 0 & 323 & 0  \\ \bottomrule
    \end{tabularx} 
    \end{subtable}  
\hfil
    \begin{subtable}{0.45\linewidth}
    \centering
    \begin{tabularx}{\textwidth}{@{} X *5{c} @{}}
\toprule
   & \multicolumn{5}{c}{\textit{pgmm}} \\ \cmidrule(r){2-6} 
Cluster & 1 & 2 & 3 & 4 & 5 \\ \midrule
Northern Italy & 0 & 0 & 88 & 63 & 0 \\
Sardinia & 0 & 0 & 0 & 0 & 98 \\ 
Southern Italy & 195 & 128 & 0 & 0 & 0\\ \bottomrule
    \end{tabularx} 
    \end{subtable}
\hfil
\vspace{0.5cm}
    \begin{subtable}{0.45\linewidth}
    \centering
    \begin{tabularx}{\textwidth}{@{} X *5{c} @{}}
\toprule
  & \multicolumn{5}{c}{\textit{fabMix}} \\ \cmidrule(r){2-6} 
Cluster & 1 & 2 & 3 & 4 & 5  \\ \midrule
Northern Italy & 60 & 0 & 0 & 0 & 91  \\
Sardinia & 0 & 98 & 0 & 0 & 0  \\ 
Southern Italy & 0 & 0 & 201 & 122 & 0 \\ \bottomrule
    \end{tabularx} 
    \end{subtable}  
  \end{center}
  \vspace{-5mm}
  \caption{Confusion matrices between the estimated and true cluster assignments to three areas of the Italian olive oils data set. The estimated cluster assignments are provided by $(MF)^2A$, IMIFA, \textit{pgmm} and \textit{fabMix} models.}
  	\label{table:olive:areas}
\end{table}

\subsection{Benchmark data: Italian olive oils} \label{olive}
	
The Italian olive oils data set (\cite{Forina1983}) has also been used in the literature for testing clustering and factor-analytical models (see, e.g. \cite{Murphy2019}) and is available in the R package \textit{FlexDir}.
The data describe the composition of 8 fatty acids in $T=572$ Italian olive oils, which originate from three areas: southern and northern Italy and Sardinia. Each area breaks down into several regions: southern Italy comprises north Apulia, Calabria, south Apulia, and Sicily; Sardinia is divided into inland and coastal Sardinia; and northern Italy comprises Umbria and east and west Liguria. Hence, one can assume that the true number of clusters should probably correspond to either $3$ areas or $9$ regions.

Table \ref{tab:olive} presents the results of applying our dynamic $(MF)^2A$ model and the other three MFA models to the Italian olive oils data. Due to a rather small number of 
$p=8$ variables  in the data set, the initial number of cluster-specific factors in the $(MF)^2A$ algorithm, which is usually set at $H = \lfloor \frac{p-1}{2} \rfloor$, was replaced by $H=p$. 
As it is unclear if the clustering should be done according to areas or regions, we calculated the ARI and the misclassification rate for both cases.

\renewcommand{\arraystretch}{1.5}
\begin{table}[t!]
  \begin{center}
    \small
    \begin{subtable}{0.45\linewidth}
    \centering
    \begin{tabularx}{\textwidth}{@{} X *5{c} @{}}
\toprule
  & \multicolumn{5}{c}{Dynamic $(MF)^2A$} \\ \cmidrule(r){2-6} 
Cluster & 1 & 2 & 3 & 4 & 5  \\ \midrule
North Apulia & 0 & 0 & 25 & 0 & 0 \\ 
South Apulia & 197 & 0 & 9 & 0 & 0 \\ 
Calabria & 0 & 0 & 56 & 0 & 0\\ 
Sicily & 0 & 0 & 36 & 0 & 0\\ 
Inland Sardinia & 0 & 0 & 0 & 0 & 65 \\ 
Coastal Sardinia & 0 & 0 & 0 & 0 & 33\\ 
Umbria & 0 & 50 & 0 & 0 & 0 \\ 
East Liguria & 0 & 40 & 0 & 10 & 0 \\ 
West Liguria & 0 & 0 & 0 & 51 & 0 \\  \bottomrule
    \end{tabularx} 
    \end{subtable}
\hfil
\vspace{0.5cm}
    \begin{subtable}{0.45\linewidth}
    \centering
    \begin{tabularx}{\textwidth}{@{} X *5{c} @{}} \toprule
  & \multicolumn{5}{c}{IMIFA} \\ \cmidrule(r){2-6} 
Cluster & 1 & 2 & 3 & 4 & 5 \\ \midrule
North Apulia & 0 & 0 & 0 & 25 & 0  \\
South Apulia & 0 & 0 & 0 & 206 & 0  \\ 
Calabria & 0 & 0 & 0 & 56 & 0   \\
Sicily & 0 & 0 & 0 & 36 & 0  \\
Inland Sardinia & 0 & 65 & 0 & 0 & 0  \\
Coastal Sardinia & 0 & 33 & 0 & 0 & 0  \\
Umbria & 0 & 0 & 47 & 0 & 3  \\
East Liguria & 0 & 0 & 0 & 0 & 50  \\
West Liguria & 48 & 0 & 3 & 0 & 0  \\ \bottomrule
    \end{tabularx} 
    \end{subtable}  
\hfil
    \begin{subtable}{0.45\linewidth}
    \centering
    \begin{tabularx}{\textwidth}{@{} X *5{c} @{}}
\toprule
   & \multicolumn{5}{c}{\textit{pgmm}} \\ \cmidrule(r){2-6} 
Cluster & 1 & 2 & 3 & 4 & 5 \\ \midrule
North Apulia & 0 & 25 & 0 & 0 & 0 \\
South Apulia & 195 & 11 & 0 & 0 & 0 \\ 
Calabria & 0 & 56 & 0 & 0 & 0 \\
Sicily & 0 & 36 & 0 & 0 & 0 \\
Inland Sardinia & 0 & 0 & 0 & 0 & 65 \\
Coastal Sardinia & 0 & 0 & 0 & 0 & 33 \\
Umbria & 0 & 0 & 50 & 0 & 0  \\
East Liguria & 0 & 0 & 37 & 13 & 0 \\
West Liguria & 0 & 0 & 0 & 51 & 0\\ \bottomrule
    \end{tabularx} 
    \end{subtable}
\hfil
\vspace{0.5cm}
    \begin{subtable}{0.45\linewidth}
    \centering
    \begin{tabularx}{\textwidth}{@{} X *5{c} @{}}
\toprule
  & \multicolumn{5}{c}{\textit{fabMix}} \\ \cmidrule(r){2-6} 
Cluster & 1 & 2 & 3 & 4 & 5  \\ \midrule
North Apulia & 0 & 0 & 0 & 25 & 0 \\
South Apulia & 0 & 0 & 197 & 9 & 0  \\ 
Calabria & 0 & 0 & 1 & 55 & 0  \\
Sicily & 0 & 0 & 3 & 33 & 0  \\
Inland Sardinia & 0 & 65 & 0 & 0 & 0  \\
Coastal Sardinia & 0 & 33 & 0 & 0 & 0  \\
Umbria & 10 & 0 & 0 & 0 & 40  \\
East Liguria & 50 & 0 & 0 & 0 & 0  \\
West Liguria & 0 & 0 & 0 & 0 & 51 \\ \bottomrule
    \end{tabularx} 
    \end{subtable}  
  \end{center}
  \vspace{-5mm}
  \caption{Confusion matrices between the estimated and true cluster assignments to nine regions of the Italian olive oils data set. The estimated cluster assignments are provided by $(MF)^2A$, IMIFA, \textit{pgmm} and \textit{fabMix} models.}
  	\label{table:olive:regions}
\end{table}	

All four models discovered $5$ clusters and all placed Sardinia to a separate cluster while the clustering assignment of northern and southern Italy differs between models (see Table \ref{table:olive:areas} and \ref{table:olive:regions} for the classification into areas and regions, respectively). Interestingly, IMIFA model delivers a better performance in clustering into big areas, while the other three models achieve a significantly better clustering result with respect to smaller regions. The major difference in the performance of the IMIFA model is that it places southern Italy into one cluster, while the other models split it into South Apulia and the rest. IMIFA also splits northern Italy into three groups, roughly corresponding the three regions. While the other three models split northern Italy into two groups, the allocation of the regions into these groups varies between models. The dynamic $(MF)^2A$ and the \textit{pgmm} models allocate Umbria and the biggest part of the East Liguria into one cluster, and the \textit{fabMix} model groups the biggest part of Umbria together with West Liguria. The ARIs of the dynamic $(MF)^2A$, \textit{pgmm} and \textit{fabMix} models are very similar to each other, with the ARI of the dynamic $(MF)^2A$ model being marginally better than the ARIs of the \textit{pgmm} and \textit{fabMix} models. 
	
	\subsection{Real data: Eurozone inflation rates} \label{EZInflation}
	
\begin{figure}[t!]
  \centering
  \includegraphics[width=0.80\textwidth, height=0.60\textwidth]{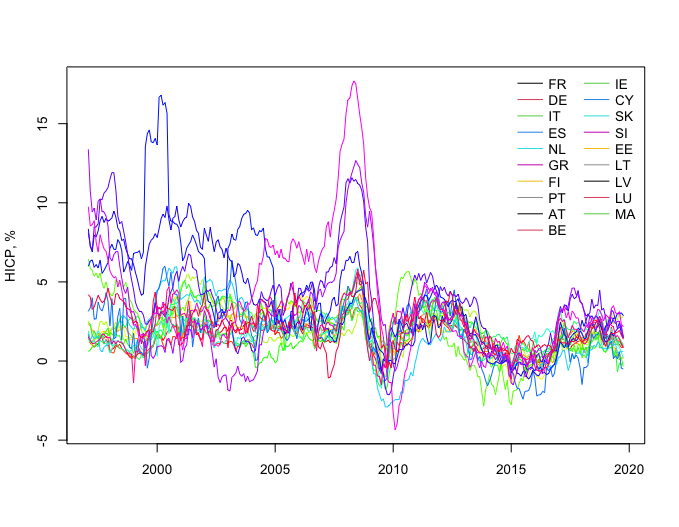}
  \vspace{-5mm}  
  \caption{HICP year-on-year inflation rates of $19$ Eurozone countries for the period February $1997$ - October $2019$.} \label{fig:infl}
\end{figure}

Next, we employ the dynamic $(MF)^2A$ model to analyse the structure of the data consisting of the Harmonised Index of Consumer Prices (HICP) inflation rates for $p=19$ Eurozone countries for the period from February $1997$ to October $2019$, which makes in total $T=273$ observation. Figure \ref{fig:infl} illustrates the path of these time series for the reported period. In this case, clustering is performed with respect to the time dimension, which seems a natural choice as in different countries factors which drive their inflation rates may differ in various time periods (for example, some countries in the data set joined the single currency area later than others). The sampler was run for $50,000$ iterations, $20$\% of which were discarded as a burn-in. The data were demeaned and unit-scaled and the hyperparameters were used as in Table \ref{tab:hyperpar}.

\renewcommand{\arraystretch}{1.3}
\begin{table}[htb!] 
  \begin{center}
    \small
    \begin{tabular}{c|c|c} 
    \hline
      Model & \# clusters & \# factors \\
      \hhline{=|=|=}
      Dynamic $(MF)^2A$ & 6 & 3,2,2,3,3,2  \\
      \hline
      IMIFA & 20 & 2,2,3,2,2,2,2,2,2,3,2,3,3,2,3,2,2,2,2,3  \\
      \hline
    \end{tabular}
  \end{center}
  \vspace{-5mm}
  \caption{Results of fitting the dynamic $(MF)^2A$ and IMIFA models on the Eurozone inflation rates data set.}
  	\label{tab:infl}
\end{table}	

\begin{figure}[t!]
  \centering
  \includegraphics[width=0.80\textwidth, height=0.60\textwidth]{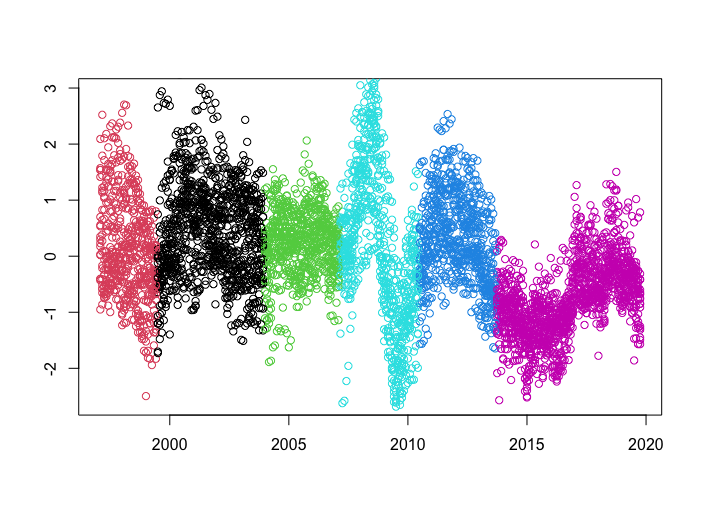}
  \caption{Cluster assignments of the Eurozone inflation rates data set according to the dynamic $(MF)^2A$ model.} 
  \vspace{-5mm}
  \label{fig:clust}
\end{figure}

The results of applying the dynamic $MF)^2A$ model to the inflation data are presented in Table \ref{tab:infl}. 
The six clusters estimated by the dynamic $(MF)^2A$ model show a clear time-related pattern as they are situated one after another on the time line, see Figure \ref{fig:clust}. 
 Cluster $1$, marked by the coral colour, contains observations in the period February $1997$ - June $1999$, which roughly corresponds to the period before the introduction of the Euro\footnote{The Euro was launched as a currency for accounting purposes and electronic payments on January $1$, $1999$ while coins and banknotes were introduced on January $1$, $2002$.}. Cluster $2$ (black) covers the period from July $1999$ till December $2003$. The $3$d cluster, depicted green in Figure \ref{fig:clust}, contains observations from January $2004$ till February $2007$, and corresponds to the period between the extension of the European Union by ten new members\footnote{On May $1$, $2004$, ten new members joined the EU, namely, Cyprus, Malta, Czechia, Estonia, Hungary, Latvia, Lithuania, Poland, Slovakia and Slovenia.} and the financial crisis. The fourth cluster, coloured light blue on the chart, covers the period between March $2007$ and June $2010$ and encompasses the financial crisis $2007-2008$ and the subsequent recession. The period from July $2010$ to September $2013$, assigned to cluster $5$ (dark blue), was marked by the European sovereign debt crisis which resulted in bailout packages for several Eurozone countries. Finally, October $2013$ - October $2019$ was a period of extremely low and at times even negative inflation rates amongst the Eurozone countries, during which the European Central Bank struggled to stimulate inflation with a very loose monetary policy.

\begin{figure}[h]
  \centering
  \includegraphics[width=0.80\textwidth, height=0.60\textwidth]{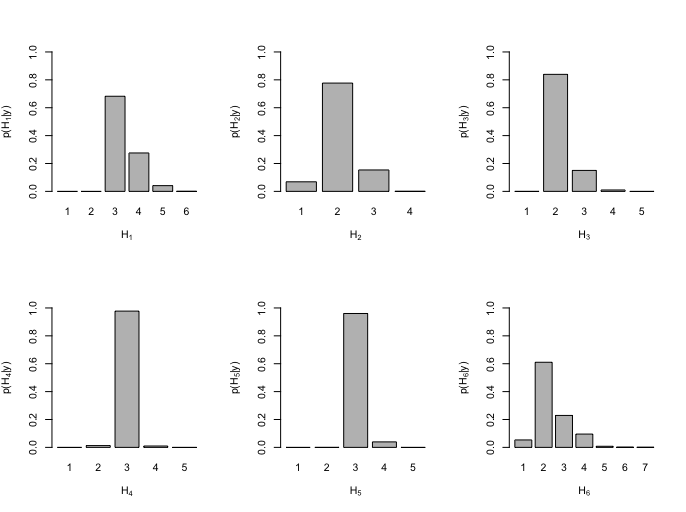}
  \caption{Posterior distribution of $H_k$ under the dynamic $(MF)^2A$ model for the Eurozone inflation rates data.} \label{fig:fact}
\end{figure}

With regard to the number of factors, 
Figure \ref{fig:fact} illustrates the posterior distribution of $H_k$ under the dynamic $(MF)^2A$ model. 
It is interesting to note that the cluster-specific number of factors estimated by the model is higher in periods marked by crises. Thus, in cluster $4$, which covers the period of the financial crisis $2007-2008$ and the following recession, the  model estimated three active factors. In the subsequent period, marked by the sovereign debt crisis, the estimated number of active factors is also three, as well as in the first period which precedes the introduction of the euro. In all other periods (clusters $2$, $3$ and $6$) the estimated number of latent factors is two.  

For comparison, we also ran the IMIFA model on the inflation data set, but not the \textit{pgmm} and \textit{fabMix}. As the main point of applying an MFA model to this type of data is to see if the data is driven by different factors in different time periods, the constraint of a common number of factors in all clusters, which is a feature of the two latter models, appears to be too restrictive. The IMIFA model estimated $20$ clusters (which is almost equal to the number of years ($21$) covered by the time period) with either $2$ or $3$ latent factors in each cluster.

\section{Conclusion} \label{Conclusion}

We proposed a novel model 
in the MFA framework which allows 
fully automatic inference on the number of non-empty components in the mixture and the number of latent factors in the cluster-specific factor-analytical models while keeping both dimensions finite at all times. 
This was done by employing the connection between nonparametric Bayesian process priors and their finite representations, connecting the MFM framework with the ESP class of priors (\cite{SFS2022}) in the factor-analytical part. This approach allowed to eliminate some of the drawbacks of the nonparametric models such as computational inefficiency and identification difficulties, which is especially important for factor-analytical models. 
Posterior inference is performed solely within Gibbs sampler steps without any adaptive mechanisms and all information is kept and stored at each iteration of the sampler. All influential parameters are learned from data, which makes it possible to use the dynamic $(MF)^2 A$ model on various data sets with no or little additional tuning. Some hyperparameter tuning may become necessary when working with data sets of essentially different nature, however, the hyperparameter values we provided proved to be rather universal and can be employed for data sets of various sizes and structure, including time series data.

Future research directions could include, for example, introducing element-wise shrinkage for the columns of cluster-specific factor loading matrices, which could help to achieve more exact identification in sparse factor models. Making hyperparameters of cluster-specific factor models, namely $\alphaB$, $b_0$ and $b_\theta$, cluster-specific could improve the model's performance in settings with differently sized clusters, where the dynamic $(MF)^2 A$ model sometimes struggled to distinguish smaller clusters. Alternatively, the triple gamma prior (\cite{Cadonna2020}) could be employed for the spike and the slab distributions instead of the inverse gamma priors (\cite{SFS2022}). This could improve mixing and uncertainty quantification of the number of cluster-specific factors.

\nocite{}
\setlength\bibitemsep{1.2\itemsep}
\printbibliography

\end{document}